\documentclass[sensors,article,accept,moreauthors,pdftex]{Definitions/mdpi} 
\makeatletter
\AtBeginDocument{\let\hl\@firstofone}
\makeatother

\firstpage{1} 

\pubvolume{00}
\issuenum{1}
\articlenumber{0}
\pubyear{2021}
\copyrightyear{2021}
\datereceived{} 
\dateaccepted{} 
\datepublished{} 
\hreflink{https://doi.org/} 

\usepackage{algpseudocode,algorithm,algorithmicx,float} 
\usepackage{graphicx, subfig}
\usepackage{gensymb}
\usepackage{booktabs}
\usepackage{tabularx}
\usepackage{csquotes}
\usepackage{amssymb}
\usepackage{pifont}
\algdef{SE}[VARIABLES]{Variables}{EndVariables}
{\algorithmicvariables}
{\algorithmicend\ \algorithmicvariables}
\algnewcommand{\algorithmicvariables}{\textbf{global variables}}

\Title{An Experimental Analysis of Attack Classification Using Machine Learning in IoT Networks}
\TitleCitation{An Experimental Analysis of Attack Classification Using Machine Learning in IoT Networks}

\Author{{\hl{Andrew Churcher} 
}$^{1}$, {Rehmat Ullah} $^{2,}$*, {Jawad Ahmad} $^{1}$, {Sadaqat \hl{Ur} 
 Rehman} $^{3}$, {Fawad Masood} $^{4}$, {Mandar Gogate} $^{1}$, Fehaid Alqahtani $^{5}$, {Boubakr Nour} $^{6}$ and William J. Buchanan $^{1}$}

\AuthorNames{Andrew Churcher, Rehmat Ullah, Jawad Ahmad, Sadaqat Ur Rehman, Fawad Masood, Mandar Gogate, Fehaid Alqahtani, Boubakr Nour and William J. Buchanan}

\AuthorCitation{Churcher, A.; Ullah, R.; Ahmad, J.; Ur Rehman, S.; Masood, F.; Gogate, M.; Alqahtani, F.; Nour, B.; Buchanan, W.J.}
\address{%
$^{1}$ \quad School of Computing, Edinburgh Napier University, Edinburgh EH10 5DT, UK; \linebreak 40344090@live.napier.ac.uk (A.C.); jawadkhattak@ieee.org (J.A.); mandar.gogate@napier.ac.uk (M.G.); B.Buchanan@napier.ac.uk (W.J.B.) \\
$^{2}$ \quad School of Electronics, Electrical Engineering and Computer Science, Queen’s University, Belfast BT9 5BN, UK\\
$^{3}$ \quad Department of Computer Science, Namal Institute, Mianwali 42250, Pakistan; sadaqat.rehman@namal.edu.pk\\
$^{4}$ \quad College of Information Engineering, Yangzhou University, Yangzhou 225127, China; fawadmasood09@mail.ist.edu.pk\\
$^{5}$ \quad Department of Computer Science, King Fahad Naval Academy, Al Jubail 35512, Saudi Arabia; f-alqahtani@rsnf.gov.sa\\
$^{6}$ \quad School of Computer Science and Technology, Beijing Institute of Technology, Beijing 100081, China; n.boubakr@bit.edu.cn}
	
	\corres{Correspondence: r.ullah@qub.ac.uk; Tel.: +44-7459-408406}

\abstract{In recent years, there has been a massive increase in the amount of Internet of Things (IoT) devices as well as the data generated by such devices. The participating devices in IoT networks can be problematic due to their resource-constrained nature, and integrating security on these devices is often overlooked. This has resulted in attackers having an increased incentive to target IoT devices. As the number of attacks possible on a network increases, it becomes more difficult for traditional intrusion detection systems (IDS) to cope with these attacks efficiently. {In this paper, we   highlight  several machine learning (ML) methods such as k-nearest neighbour (KNN), support vector machine (SVM), decision tree (DT), naive Bayes (NB), random forest (RF), artificial neural network (ANN), and logistic regression (LR) that can be used in IDS. In this work, ML algorithms are compared for both binary and multi-class classification on Bot-IoT dataset. Based on several parameters such as accuracy, precision, recall, F1 score, and log loss, we   experimentally compared the aforementioned ML algorithms. In the case of HTTP distributed denial-of-service (DDoS) attack, the accuracy of RF is 99\%. Furthermore, other simulation results-based precision, recall, F1 score, and log loss metric  reveal that RF outperforms on all types of attacks in binary classification. However, in multi-class classification, KNN outperforms other ML algorithms with an accuracy of 99\%, which is 4\% higher than RF. }}

\keyword{Internet of Things (IoT); IoT attacks; security; intrusion detection systems; privacy; machine learning; ML models; multi-class classification}

\begin{document}

\setcounter{section}{0} 

\section{Introduction}
The Internet of Things (IoT) offers a vision where devices with the help of sensors can understand the context  and through networking functions can connect with each other \cite{dorsemainegaulierwarykheirurien2015}. The devices in the IoT network can be employed for collecting information based on the use cases. These include retail, healthcare, and manufacturing industries that use IoT devices for tasks such as tracking purchased items, remote patient monitoring,  and fully autonomous warehouses. It is reported that the amount of IoT devices has been growing every year with the predicted amount of devices by 2025 reaching 75.44 billion \cite{statistaresearchdepartment2019}. Such a massive surge of IoT devices ultimately results in more attackers to target IoT networks. Reports state that most of the attack traffic generated on IoT networks is automated through various means such as scripts and malware \cite{doffman2019}. The increase in attacks combined with the autonomous nature of the attacks is a problem for IoT networks as the devices are mostly used in a fire and forget fashion for years without any human interaction. This combined with the limitations of IoT devices including limited processing power and bandwidth means that providing adequate security can be difficult, which can result in network layer attacks such as denial of service (DoS). Therefore, it is important to research ways to identify this kind of traffic on networks which can be used in intrusion detection and prevention systems. 

Machine learning (ML) methods can be exploited to detect malicious traffic in intrusion detection and prevention systems. ML is a subset of artificial intelligence (AI) that involves using algorithms to learn from data and make predictions based on the data provided \cite{furbush2018}. 
ML has many applications including in  retail, healthcare, and finance where AI algorithms may be applied  for predicting customer spending habits, predicting medical problems in patients, and detecting bank fraud, respectively \cite{jmj2018}.

Due to the large yearly increases in cyberattacks that are being seen on a yearly basis, ML methods are being incorporated to help tackle the increasing threats of cyberattacks. ML has several uses within the field of cybersecurity, such as network threat analysis, which can be defined as the act of analyzing threats to the network \cite{dosal2018}. ML can be beneficial in this task as it is able to monitor incoming and outgoing traffic to identify potentially suspicious traffic \cite{groopman2019}. This area of research is known as intrusion detection and is a widely known research area. ML can be applied to intrusion detection systems (IDS) to help improve the systems ability to run autonomously and increase the accuracy of the system when raising the alarm on a suspected attack \cite{cuelogictechnologies2019}. To this end, our primary role is to identify the best ML methods for detecting attacks on IoT networks, using a state-of-the-art dataset by utilizing both binary and multi-class classification testing.

{The main contributions of this paper can be summarized as follows:}
{
\begin{enumerate}
	\item We conduct an in-depth and comprehensive survey on the role of various ML methods and attack detection specifically in regards to IoT networks.
	\item We evaluate and compare the state-of-the-art ML algorithms in terms of various performance metrics such as confusion matrix, accuracy, precision, recall, F1 score, log loss, ROC AUC, and Cohen's kappa coefficient (CKC).	
	\item We evaluate the results comparing binary class testing as well as examining the results of the multi-class testing. 
\end{enumerate} 
}

{The rest of the paper is organized as follows:
Table \ref{acronyms} lists all the  abbreviations  used in the paper. Section \ref{section2} is devoted to a literature review involving investigating IoT intrusion detection techniques as well as ML methods and how they are being used to aid intrusion detection efforts specifically in regards to IoT networks. Details of various attacks that can occur in IoT networks are also showcased with an explanation of how the various ML methods and performance metrics work. Section \ref{section3} explains the performance evaluation,  which also includes an in-depth examination of the data used in the datasets. The models are compared against each other for both binary and multi-class classification with an overall best model being selected. Finally, Section \ref{section4} draws a conclusion.}

\begin{specialtable}[H]
	\caption{{Abbreviations  and their explanations.}\label{acronyms}}
\setlength{\tabcolsep}{2.4mm}
	
	\begin{tabular}{cccc}
		\toprule 
		\textbf{Acronym} & \textbf{Explanation} & \textbf{Acronym} & \textbf{Explanation}\\
		\midrule 
		IDS & Intrusion Detection Systems & ANN & Artificial Neural Network\\
		ML & Machine Learning & KNN & K-nearest Neighbour\\
		SVM & Support Vector Machine & DT & Decision Tree\\
		NB & Naive Bayes & RF & Random Forest\\
		LR & Logistic Regression & DDoS & Distributed Denial-of-Service\\
		IoT & Internet of Things & CKC &Cohen's Kappa Coefficient\\
		TP & True Positive & TN & True Negative\\
		FP & False Positive & FN & False Negative\\
	TPR & True Positive Rate & FPR & False Positive Rate\\
		\bottomrule
	\end{tabular}
\end{specialtable}
\section{Background and Related Work}
\label{section2}

This section presents the background and examines current literature that would clear up the picture for the reader about the design of the experiments conducted in this paper. Firstly, we discuss IDS including the use of ML used in attack detection and the related work which would help with selecting the algorithms to be used as well as identifying any datasets that could be utilized for testing the models. Each algorithm is explored with further research into the suitability of the algorithm for use in an IDS. The IoT is also described including the attacks that are used in the dataset that has been selected.

\subsection{Intrusion Detection System}

An IDS is a tool that allows a network to be monitored for potentially harmful traffic. An IDS can be implemented using two distinct types: signature-based detection and anomaly-based detection. A signature-based IDS uses a database of existing attack signatures and compares the incoming traffic with the database, meaning that an attack can be detected only if the signature is already available in the database. An anomaly-based IDS monitors network traffic and attempts to identify any traffic that is abnormal in regards to the normal network traffic.

The signature-based detection approach has a major flaw as a signature-based IDS will always be susceptible to a zero-day attack or an attacker that modifies the attack to hide from the signature database.   Anomaly-based IDS are much better suited to use ML as the IDS can be trained to detect the difference between normal traffic and attack traffic.
However, integrating ML with IDS  is not a silver bullet and may result in some problems. Research conducted by Sommer    and    Paxson \cite{5504793} identified several problems where one important problem is that models can produce false positives, which can render the IDS unusable due to normal data causing the IDS to alert the system. Even though the research is very outdated, this is still a major problem when using ML with IDS. As a result of this, it is of paramount importance to identify models that produce the lowest number  of false positives.

\subsection{IoT Intrusion Detection Using Machine Learning}

ML is a subset of AI that involves giving an algorithm or in this case a model  a dataset  which will be used to identify patterns that can be used to make predictions with future data. 
There has been limited research devoted to IDS using ML on IoT networks. To this end, recently a study used the Defense Advanced Research Projects Agency (DARPA) ML datasets to test the models such as support vector machine (SVM), Naive Bayes (NB), random forest (RF), and multi-layer perceptron \cite{foleymoradpoorochenyi2020}. The results of this research were presented in terms of root mean squared error, mean absolute percentage error, receiver operating characteristic curve, and accuracy, yielding good results with RF being one of the top models. However, this research has two main limitations: Firstly, it used the DARPA datasets, which were over 20 years old at the time of writing. Secondly, it  was not performed for multi-class testing using the datasets. 

The research was also conducted using the Bot-IoT dataset that used the models k-nearest neighbour (KNN), quadratic discriminant analysis, iterative dichotomiser 3, RF, adaptive boosting, multi-layer perceptron, and NB \cite{alsamirialsubhi2019}. The research did yield very good results in terms of accuracy, precision, recall, F1 score, and time. This study used an up-to-date dataset as well as a wide variety of ML models. However, this research did not include any multi-class testing for any of the models. 

In regards to multi-class classification, the authors  of \cite{hasanislamzarifhashem2019} used several ML methods. This research compared the algorithms such as logistic regression (LR), decision tree (DT), RF, and artificial neural network (ANN) using a dataset created by the researchers which was not available for public use. It was concluded in the study that RF was the best model for multi-class classification. This research shows that with multi-class classification it is possible to achieve high results. Testing with additional algorithms could help bolster the results of the research.

Overall, there is currently a lack of research into intrusion detection within the area of IoT networks. This could be due to the lack of datasets as well as lack of real hardware with all datasets being comprised of simulated IoT devices on regular computers. There is also a lack of research into multi-class classification,  which  could be due to the lack of a dedicated multi-class dataset. With all available datasets being created with binary classification in mind, performing multi-class testing requires the datasets to be merged into one with proper labelling for each class.

Various ML models can be utilized to perform ML tasks, each with their own mathematical equations powering the analysis of the data presented. In the next subsections, we discuss various ML algorithms for our analysis such as: (i) KNN; (ii) SVM; (iii) DT; (iv) NB; (v) LR; and (vi) ANN.

\subsubsection{K-Nearest Neighbor}

KNN is a supervised learning model that is considered to be one of the simplest ML models available \cite{brownlee2019}. KNN is referred to as a lazy learner  because there is no training done with KNN; instead, the training data are  used when making predictions to classify the data \cite{brownlee2019}. KNN operates under the assumption that similar data points will group and finds the closest data points using the K value, which can be set to any number \cite{harrison2019}. KNN is a suitable model to be used for intrusions detection as showcased with several pieces of research conducted. The authors   of \cite{liaovemuri2002} examined the effectiveness of KNN at distinguishing between attack and normal data. The results of this research show  that KNN was an effective model of detecting attack data and had a low false-positive rate. Moreover, recent research also examined the effectiveness of KNN \cite{nikhithajabbar2019} with a similar consensus being met. The research showed that KNN was an effective model beating SVM and DT.

\subsubsection{Support Vector Machine}
{ Support Vector Machine (SVM) is a supervised learning algorithm that uses a hyperplane to separate the training data to classify future predictions. The hyperplanes divide a dataset into two classes and they are decision boundaries that help classify the data points. A hyperplane can be represented as a line or a plane in a multi-dimensional space} and is used to separate the data based on the class they belong to. It does this by finding the maximum margin space between the support vectors. SVM is a suitable model for intrusion detection as evident by the large amount of research conducted over the years. One older piece of research created an enhanced SVM model for intrusion detection \cite{yaozhaofan2006}. The research was successful at creating the model but proved to be only a slight improvement over regular SVM, showing that the model even without enhancements or augmenting is capable of accurately classifying attack data. Other more recent research compared SVM and ANN's ability to classify attack data \cite{cahyohidayatadhipta2016}. 
As previously mentioned, SVM relies  on placing a hyperplane to separate data which can be expressed as follows:

\begin{equation}a x+b=0\end{equation}
where $a$ is the vector of the same dimensions as the input feature vector $x$ and $b$ is the bias. In this case, $ax$ can be written as \(a^1 x^1 + a^2 x^2 +...+ a^n x^n\) where $n$ is the number of dimensions of the feature vector $x$. When making predictions, the following expression is used:

\begin{equation} y = sign(ax - b)\end{equation}
where $sign$ is a function that returns either $+1$ or $-1$ depending if the input is a positive number or a negative number respectively. This value is used to determine the prediction of what class the feature vector belongs to. $x_i$ is the feature vector and $i$ and $y_i$ is the label that can either be $+ 1$ or $-1$ and can be written as the follows:

\[ ax_i - b \geq + 1\; if y_i = +1\]
\[ ax_i - b \leq - 1\; if y_i = -1\]

{SVMs use kernels and kernel is basically a set of mathematical functions. The kernel is used to take data as an input and transform them  into the required form of processing data. The kernels can be linear, nonlinear, polynomial, Gaussian kernel, Radial basis function (RBF), sigmoid, etc. }

\subsubsection{Decision Tree}

DT is a supervised learning algorithm that is useful to present a visual representation of the model. A DT uses a hierarchical model that resembles a flow chart which has several connected nodes. These nodes represent tests on the attribute in the dataset with a branch that leads to either another node or a decision on the data being classified \cite{sharmakumar2016}. The training data are  used to build the tree with the prediction data being run through the nodes until the data can be classified. DT is a suitable model for intrusion detection based on the research conducted. One fairly recent piece of research compared DT with several other models including NB and KNN \cite{stamper2015}. The results show  that DT was one of the better models along with NB when compared to ANN's which dominate IDS research. Other research created an IDS for connected vehicles in smart cities \cite{aloqaily2019}. This research showed that the model that used DT was the best model with high accuracy and a low false positive rate.
As previously mentioned, DT creates a hierarchical model using the training data to create nodes that act as tests for making predictions. {When making DT, the root node needs to be selected as well as selecting the nodes that make up the DT}. In this regard, there are many ways to do this with entropy being used in this case. Entropy is used to measure the probability of a data point being incorrectly classified when randomly chosen and is expressed as follows:
{
\begin{equation}
	E = \sum^{c}_{i=1}-p_{i}\; log_{2} \ (p_{i})
\end{equation}
where $p_{i}$ is the probability of the data being classified to a given class of $i$ and $c$ is the number of classes. The attribute with the lowest entropy would be used for the root node. }

\subsubsection{Random Forest}

RF is a supervised learning algorithm that is seen to be an improvement on the DT model. The random aspect of the model comes from two key concepts. The first is that,  when training the model, each tree is given a random assortment of the data which can result in some trees using the same data multiple times. The reason behind this is to lower the variance of the model, which lowers the difference in the predicted results scores \cite{koehrsen2018}. The second concept involves only using small subset of the features when splitting the nodes in the trees \cite{dubey2018}. This is done to prevent {overfitting} when the model uses the training data to inflate the predictions made by the model \cite{brownlee2019}. When making predictions with RF, the average of each of the trees predictions is used to determine the overall class of the data;  this process is called bootstrap aggregating \cite{brownlee2019}. The reason   RF is seen as an improvement on DT is that,  instead of relying on one tree to make the classification, multiple trees with different training data and with a different selection of features are used for giving predictions. This allows for a fairer analysis of the data when making predictions. RF is proven to be a suitable model for intrusion detection. To this end, the authors  of  \cite{farnaazjabbar2016} compared RF to other frameworks used in intrusion detection. They  found that the RF model {outperformed} the other frameworks with increased accuracy, precision, recall,  and F1 score.

\subsubsection{Naive Bayes}

NB is a probabilistic algorithm that works by getting the probability of all the feature vectors and their outcome. The algorithm is used to determine the probability of an event occurring based on previous events occurring which is called posterior probability and is expressed as follows:

\begin{equation}
P(A | B) = \frac{P(B|A)P(A)}{P(B)} 
\end{equation}
where \(P(A | B)\) is the posterior probability, \(P(A)\) is known as the prior probability, \(P(B)\) is marginal likelihood (evidence), and \(P(B | A)\) is referred to as the likelihood. This formula can be applied to datasets in the following way:

\begin{equation}
P(y | x) = \frac{P(x|y)P(y)}{P(x)}
\end{equation}
where $y$ is the class variable and $x$ is the feature vector of size $n$ shown as the following:
{
\begin{equation}
 x = (x_{1},x_{2},x_{3},...,x_{n})
\end{equation} }

\subsubsection{ANN}
{An} ANN refers to a model of performing machine learning that is based on how the human brain operates and can be used to perform supervised learning. {An} ANN consists of neurons or nodes that make up the layers of the network \cite{saritas2019performance}. The three types of layers in an ANN are input, hidden, and output layers { where the input layer takes information provided and passes it onto the hidden layer. The hidden layer performs computations and transfers the data to the output layer. The output layer also performs computations and presents the output of the ANN \cite{ujjwalkarn2016}.
When performing supervised learning,  {the network is given }the inputs and expected outputs for training. The connections between the nodes in the network have numbers assigned to them called weights. When an error is made by the network, the data  are  propagated back through the network and the weights are adjusted. {This process occurs repeatedly until the error is minimized, and then the test data can be fed through the network} \cite{maind2014research}. Training an ANN is described as follows: }

{The first step in training the ANN involves multiplying the input values $x_{i}$ and the weights $w_{i}$,  and then summing the values expressed as the following:
\begin{equation}
x_{i}\cdot w_{i} = (x_{1} \cdot w_{1}) + (x_{2} \cdot w_{2}) +...+ (x_{n}\cdot w_{n}) 
\end{equation}}

The second step involves adding the summed values to the bias $b$ of the hidden layer node as expressed as the following:
{
\begin{equation}
z = x_{i}\cdot w_{i} + b 
\end{equation}}

The third step is to pass the $z$ value through an activation function such as ReLU and Softmax. ReLU $R(z)$ can be defined as follows:
\begin{equation}
\hat y = R(z) = max(0, z), 
\end{equation}
{where $z$ is the input to a neuron. When the $z$ is smaller than zero, the function will output zero, and, when the $z$ is greater or equal to zero, the output is simply the input.
} Softmax can be defined as follows:
{
\begin{equation}
\hat y = s(z)_{i} = \frac{e^{z_{i}}}{\sum\nolimits_{j=1}^{n} e^{z_{j}}} 
\end{equation}
where $e$ is the base of the \textbf{\hl{natural logarithm}}, $z$ is a vector of the inputs, and $_{i}$ and $_{j}$ indexes the input and output units, respectively.}

   To train the ANN, the loss needs to be calculated so the network can effectively evaluate its performance and make the appropriate changes. Once the loss has been calculated, the next step is to minimize this loss by changing the weights and the biases. Knowing how the cost function $C$ {(which is is a measure of ``how good'' a neural network did with respect to its given training sample and the expected output)} changes in relation to weights $w_i$ can be done using gradients. Using the following chain rule, the gradient of the cost function in relation to the weights can be calculated:

\begin{equation}
\frac{\partial C}{\partial w_i} = \frac{\partial C}{\partial \hat y} \times \frac{\partial \hat y }{\partial z} \times \frac{\partial z}{\partial w_i} 
\end{equation}
where $\frac{\partial C}{\partial \hat y}$ is the gradient of the cost function, $\frac{\partial \hat y }{\partial z}$ is the gradient of the predicted value,  and $\frac{\partial z}{\partial w_i}$ is the gradient of $z$ in regards to $w_i$.

ANN is the most suitable model for IoT attacks detection and {has had} many implementations. Recently, the authors  of  \cite{anithaarockiam2019} implemented an ANN based model for detecting IoT based {attacks.} The model was successful and can be used on IoT networks to perform intrusion detection. In \cite{shenfielddayayesh2018}, the  implementation is done for intrusion detection using ANNs. This research had very good results with the model having a near perfect accuracy and a very low false positive rate.

\subsubsection{Logistic Regression}

LR is a supervised learning algorithm that uses the logistic function also known as the Sigmoid function. Logistic regression is similar to linear regression except, instead of predicting data that  are  continuous, it is used for classifying data either true or false. Linear regression can have any value, whereas LR has values between 0 and 1 \cite{rajput2018}. Logistic regression is a model that is less represented in intrusion detection than other models. Its suitability for use in intrusion detection is not as well established as the previous models. However, some research has examined a logistic regression based intrusion detection model \cite{7060117}. This model was tested using multi-class classification and was able to outperform the other models.

As previously mentioned, logistic regression can be thought of as linear regression but for classification problems. The reason that logistic regression is used is because with linear regression the hypothesis {$h_{o}(x)$} can be greater than one or less than zero. With logistic regression, the hypothesis is between zero and one, e.g., {$0 \leq h_{o}(x) \leq 1$, where $h_{o}$ is a single hypothesis  that maps inputs to outputs and can be evaluated and used to make predictions.}

   To get a value between zero and one, the Sigmoid function is used which is represented as follows:
{
\begin{equation}
S(x) = \frac{1}{1 + e^{-x}} 
\end{equation}}

This function returns a number between 0 and 1 which can be mapped to a particular class of data by using a decision boundary to determine the likelihood of the data of a certain class, which can be expressed as follows:

\[ p \geq 0.5\ class = 1\]
\[ p < 0.5 \ class = 0\]

 Once the threshold is set, predictions can be made using the Sigmoid function to determine the likelihood that the data belongs to class 1 as follows:

\begin{equation}
S(class = 1) = \frac{1}{1 + e^{-x}}
\end{equation}

This function gives back a number that represents the probability that the data should be classified as Class 1. With the previously defined threshold, if the number is 0.5 or above then the data will be classified as Class 1, and anything less than 0.5 will be classified as class 0. 

The following subsection provides some details on IoT including the attacks that are used in the dataset for this paper. 

\subsection{Internet of Things Attacks}

As previously discussed, IoT is considered as a network of devices/objects communicating through wired
or wireless communication technologies \cite{9060970}. The protocols used by IoT devices are designed to be used on devices with limited computation, storage, and communication capabilities that need to conserve as much battery power as possible. Such protocols include ZigBee, radio-frequency identification (RFID), and smart Bluetooth. The relatively quick increase in IoT devices being used has resulted in a lack of standardization activities which have seen a massive influx of unsecured devices being connected to networks \cite{saleemhammoudehrazaadebisiande2018}. This in turn creates a massive attack vector allowing for a massive amount of vulnerable devices open to be exploited by attackers. 
In the following subsections, we provide relevant threats and attacks faced by IoT.
\subsubsection{Data Exfiltration}

A data exfiltration attack involves attackers gaining access to a private network and stealing data stored on the network \cite{ullahedwardsramdhanychitchyanbabarrashid2017}. This type of attack can result in the theft of data such as credit card information and personal data. Several studies have been  conducted in the field of detecting data exfiltration attacks using methods such as partially observable Markov decision process \cite{carthysinhatambemanadhata2016} and {a method that involves} capturing metadata at the file system level \cite{fadolalkarimbertino2019}.

\subsubsection{DoS and DDoS}

Denial of service (DoS) and distributed denial of service (DDoS) attacks are very similar in execution. The primary difference involves the scale of the attack. A DoS attack involves a single system and Internet connection being used to attack the victim, whereas a DDoS attack involves multiple systems and Internet connections on a global scale being used to attack the victim, which  are typically referred to as botnets \cite{maliksingh2015}. 

There are many different ways to perform either of these attacks depending on what protocol is used in the attack. These different methods include HTTP flood, TCP SYN,  and UDP flood attack, as identified by \citet{mahjabinxiaosunjiang2017}. An HTTP flood attack involves altering either the GET or POST requests sent via HTTP. A GET request is used when a client wishes to receive information from the server, whereas a POST request is used to send information to the sever such as uploading a file. Sending thousands of these requests to a server or cluster of servers at once increases the workload at the server(s) side exponentially, slowing  the entire system down or preventing  legitimate users from accessing the server(s). 

A TCP SYN attack exploits the three way handshake that occurs during a TCP connection which involves sending a SYN packet which elicits a response from the server with a SYN and ACK packet. During the attack, the destination address sent in the SYN packet is false. As a result, the server sends out SYN and ACK messages repeatedly. This process stores entries in the server's connection tables which then becomes full and prevents legitimate users from accessing the server. A UDP flood attack involves sending UDP packets with a port number and sometime a spoofed IP address as well. Once the server receives this packet,  it will check for any applications using the port in the UDP packet. The server checks for applications associated with these UDP packets and, if not found, the server sends back a ``Destination Unreachable'' packet. As more and more packets are received, the system becomes unresponsive to other clients.

Moreover, attackers are able to turn on devices such as webcams and digital video recorders (DVRs). One such example of this was the  Mirai botnet in 2016 which was able to make use of up to 400,000 devices and take down large websites such as Twitter and GitHub \cite{koliaskambourakisstavrouvoas2017}. Due to lack of security on IoT devices, paramount research has  been conducted into detecting DoS and DDoS traffic \cite{galeanobra2020,ulbinasifullah2018}. However, all such algorithms lacks the use of ML techniques. 

\subsubsection{Keylogging}
The basic function of a keylogger is to store the keystrokes made by a user on their keyboard. Keyloggers can be both hardware and software based \cite{tom2008,abukar2014}. Software keylogging is typically done by installing malware on the victim machine that saves the key strokes and relays this to the attacker. Some research has  been devoted to keylogging detection methods  (see, e.g., \cite{ortolanigiuffridacrispo2010,wajahatimranlatifnazirbilal2019}).

\subsubsection{OS Scan    and    Service Scan}

Operating system (OS) and service scans are similar in nature and can be grouped into the attack category of probing. This can be done either passively, in which the attacker gathers packets from the network, or actively, in which the attacker sends traffic and recording the responses. Since passive scanning generates no traffic, active scanning is needed for traffic to test. OS scans involve the attacker being able to discover the OS being used by the victim machine. This information can help an attacker identify the type of device, e.g., server, computer or IoT device. It can also help the attacker identify the version of the OS being used. This can help the attacker find vulnerabilities related to the OS.

There has been plethora of research conducted into using OS scans to identify if a device is an IoT device. One study used neural networks to identify if the device scanned was an IoT device \cite{yanglisun2019}. Another study used deep learning techniques to identify Raspberry Pi devices that were acting as IoT devices \cite{anejaanejaislam2018}. Both   studies show that it is possible to identify IoT devices using OS scanning techniques. 

Service scans,  more commonly referred to as port scans, involve the attacker probing a network in order to identify open ports on the network \cite{bhuyanbhattacharyyakalita2011}. This is commonly used by an attacker to gain a better insight into the types of activity on the network as well as showcasing any open ports that are vulnerable to being exploited. A port scan works by having the software used send a request to a port on another network to set up a connection. The software will then wait for a response from the network.

Due to the fact that IoT devices can range from printers to heating controllers, the ports that can be used by devices can vary. To this end, the authors  of  \cite{markowskymarkowsky2015} conducted a study performing a scan on printers to identify vulnerable ports. The results showcase  that port 9100 was a commonly opened port on printers. The port is used to carry data to and from printers over TCP. It was also noted that   gaining  access to the network using this port was a simple process. 

Port scanning can also be used to identify if a device is an IoT device. An analysis by \citet{sivanathangharakheilisivaraman2018} showed that by scanning for a small number of TCP ports it could be determined whether a device was an IoT device including information on the device itself, such as identifying a device as an HP printer. Since IoT devices are generally more vulnerable than other devices, this could be used to identify an entry point to a network. A study using an approach based on Dempster--Shafer evidence theory produced a solid groundwork for detecting port scan traffic \cite{shaochenyinye2016}. Another study proposed a new evaluation metric for IDS, which was reported to take  less time to identify port scan data than previous metrics \cite{lopezvizcainonovoafernandezcarneirocacheda2019}. Neither of these studies included IoT devices, and there is currently a lack of research into OS scans in regards to IoT devices.

{  Recently,  several efforts have been devoted for ML in IoT network \cite{9060970,rashid2020cyberattacks, soe2020machine, 8804727}. However, in most of the existing works, the performance are checked for specific types of ML algorithms, such as ANN, J48 DT, and NB without detailed performance evaluation. Although some work is based on various ML algorithms such as LR, SVM, DT, RF, ANN, and KNN, most of them are used to mitigate IoT cybersecurity threats in special environments such a smart city. Contrary to existing works, our study provides a comprehensive evaluation for both real attack and simulated attack data that were  created by simulating a realistic network at the University of New South Wales where real attacks on IoT networks  were  recorded. }

\section{Performance Evaluation}
\label{section3}
\subsection{Benchmark Data}

Our evaluation involves using several datasets with several ML models to identify the best model for correctly classifying IoT attack data. {When selecting the datasets, the two most important factors were the amount of variety in the attack data and how up-to-date the datasets are}. The datasets chosen were the bot-IoT datasets \cite{koroniotismoustafasitnikovaturnbull2019} because {they met} the two criteria previously mentioned.

\subsection{Performance Evaluation Metrics}
For evaluation, we consider the following metrics.
\subsubsection{Confusion Matrix}

A confusion matrix shows the predictions made by the model. It is designed to show where the model has correctly and incorrectly classified the data. 

The confusion matrix for binary and multi-class classification {is different}. With binary classification, the matrix shows the true positive (TP), true negative (TN), false positive (FP), and false negative (FN) results, as shown in Table \ref{CME}. The columns represent the correct classification of the data and the rows represent the available classifications.

\begin{specialtable}[H]
	\caption{Confusion matrix example.}\label{CME}
\setlength{\cellWidtha}{\columnwidth/3-2\tabcolsep-0in}
\setlength{\cellWidthb}{\columnwidth/3-2\tabcolsep-0in}
\setlength{\cellWidthc}{\columnwidth/3-2\tabcolsep-0in}
\scalebox{1}[1]{\begin{tabularx}{\columnwidth}{>{\PreserveBackslash\centering}m{\cellWidtha}>{\PreserveBackslash\centering}m{\cellWidthb}>{\PreserveBackslash\centering}m{\cellWidthc}}
		\toprule
		& \textbf{Actual Label} &\\
				\midrule
		Predicted label & No attack &Attack\\
		\midrule
		No attack & True negative & {False negative}\\
		Attack & {False positive} & True positive\\
		\bottomrule
	\end{tabularx}}
\end{specialtable}

TP and TN are when the data  are  correctly classified as either attack or no attack. FP and FN are when data  are  incorrectly predicted as the other class. When using a confusion matrix for multi-class problems, the same principles apply. However, the matrix shows all the classes which allow for observing where the mis-classification is occurring in the classes,  as shown in Table \ref{MCMME}.

\begin{specialtable}[H]
	\caption{Multi-class confusion matrix example.}\label{MCMME}
\setlength{\tabcolsep}{8.8mm}
	\begin{tabular}{cccc}
			\toprule
		& \textbf{Actual Label} &&\\
			\midrule
		Predicted label & Class 1 &Class 2&Class 3\\
	\midrule
		Class 1 & C &W&W\\
		Class 2 &W &C&W \\
		Class 3&W &W&C \\
				\bottomrule
	\end{tabular}
\end{specialtable}

In   Table \ref{MCMME}, C represents where the correct classifications are located and W represents incorrect classifications. It is to be noted that correct classifications create a diagonal path through the table from the top left corner to the bottom right corner.

\subsubsection{Accuracy}
\hl{Accuracy} 
 is a metric that can be used to identify the percentage of predictions that were classified correctly and is expressed as follows:

\begin{equation}
Accuracy = \frac{\text{Number of correct predictions}}{\text{Total number of predictions}} 
\end{equation}

This can be expanded upon by utilizing the results of a confusion matrix including TP, TN, FP, and FN and can be defined as follows:

\begin{equation}
Accuracy = \frac{\text{TP + TN}}{\text{TP + TN + FP + FN}} 
\end{equation}

\subsubsection{Precision}
\hl{Precision} 
is used to determine the ratio of correctly predicted positive outcomes against the total number of predicted positive outcomes and can be defined as follows:

\begin{equation}
Precision = \frac{\text{TP}}{\text{TP + FP}}
\end{equation}

\subsubsection{Recall}
\hl{Recall} 
 is used to determine the ratio of correctly predicted positive outcomes to all the outcomes in the given class and can be defined as follows:

\begin{equation}
Recall = \frac{\text{TP}}{\text{TP + FN}}
\end{equation}

\subsubsection{F1 Score}
\hl{F1 score} 
is the weighted average of both precision and recall which produces a number between 0 and 1. F1 score is seen as a better performance metric than accuracy and can be defined as follows:

\begin{equation}
F1 score = \frac{\text{2 $\times$ (recall $\times$ precision)}}{\text{recall + precision}}
\end{equation}

{It is to be noted that selection of \hl{F1} score or accuracy is dependent on how the data  are  distributed. The \hl{F}1 score seems a better performance metric than accuracy in the  case where the classes are highly unbalanced. \hl{F}1 score takes into account how the data  are  distributed, and, in most real-life classification problems, imbalanced class distribution exists and thus \hl{F}1 score is a better metric to be used.   Accuracy is used when the class distribution is similar and it does not take into account how the data  are distributed, which may lead to wrong conclusion. }

\subsubsection{Log Loss}

Log loss is used to measure the performance of a model by using the probability of the expected outcome. The higher   the probability of the actual class is, the higher the log loss will be. The lower score indicates that the model has performed better.

 For binary classification where number of possible classes (\emph{M}) = 2, log loss can be expressed as follows:

\begin{equation}
-{(y_i\log(p_i) + (1 - y_i)\log(1 - p_i))} 
\end{equation}

For multi-class classification where \emph{M} $>$ 2, sa eparate loss for each class label is calculated, and the results are summed, which is expressed as follows.

\begin{equation}
-\sum_{c=1}^My_{o,c}\log(p_{o,c})
\end{equation}
where \emph{M} is the number of possible classes (0, 1, 2), log is the natural logarithm, $y_i$ is a binary indicator of whether class label $i$ is the correct classification for observations, and
$p_i$ is the models prediction probability.

\subsubsection{ROC AUC}

ROC is a graph used to plot the results of the model at various thresholds when making predictions. The graph uses the true positive rate (\hl{TPR}) and false positive rates (\hl{FPR}), which are expressed as follows: 

\begin{equation}
TPR = \frac{\text{TP}}{\text{TP + FN}}
\end{equation}

\begin{equation}
FPR = \frac{\text{FP}}{\text{FP + TN}}
\end{equation}

\subsubsection{Cohen's Kappa Coefficient}

{Cohen's kappa coefficient (CKC)}, also referred to as the kappa statistic, is used to test the inter rater reliability of prediction and can be expressed as follows:

\begin{equation}
k = \frac{\text{Pr(a) \hl{-} Pr(e)}}{\text{1 \hl{-} Pr(e)}} 
\end{equation}

\noindent where Pr(a) is the observed agreement and Pr(e) is the expected agreement. This metric is useful as it compares the model against a model that guesses based on the frequency of the classes. This allows for the disparity in a dataset to be evaluated particularly with multi-class testing as the dataset has varying numbers of data points per attack. 

\subsection{Dataset Description}

The dataset named   Bot-IoT was submitted to the IEEE website on \hl{16/10/19 }and was created by the University of New South Wales (UNSW). The dataset consists of ten CSV files containing records for the following attacks on IoT networks: (i) Data exfiltration; (ii) DoS HTTP; (iii) DoS TCP; (iv) DoS UDP; (v) DDoS HTTP; (vi) DDoS TCP; (vii) DDoS UDP; (viii) Keylogging; (ix) OS Scan; and  (x) Service Scan. 
 The dataset comprises   both real attack and simulated attack data and was created by simulating a realistic network at the UNSW \cite{koroniotismoustafasitnikovaturnbull2019}.

Table \ref{DSF} shows the features used in the experiments. There are  35 columns in the dataset. However, only the ones in Table \ref{DSF} were used. When deciding what features to use, the contents of the columns are examined and any columns that have no values are removed as well as columns that contain text and columns that are deemed to be irrelevant to the overall classification of the data.

\begin{specialtable}[H] 		
	\caption{Dataset features and description.}\label{DSF}
	\setlength{\tabcolsep}{15mm}
	\begin{tabular}{l l}
		\toprule 
		\textbf{Features} & \textbf{Description}\\
		\midrule
		Stime & Record start time\\
		
		Sport & Port that data is being sent from\\
		
		Dport & Port that data is being received from\\ 
		
		Pkts & Total number of packets transferred\\
		
		Bytes & Total number of bytes transferred\\
		
		Ltime & Record last time\\
		
		Seq & Sequence number\\
		
		Dur & Record total duration\\
		
		Mean & Average duration of aggregated records\\
		
		Sum & Total duration of aggregated records\\
		
		Min & Minimum duration of aggregated records \\
		
		Max & Maximum duration of aggregated records\\
		
		Spkts & Source to destination packet count \\
		
		Dpkts & Destination to source packet count \\
		
		Sbytes & Source to destination byte count \\
		
		Dbytes & Destination to source byte count \\
		
		Rate & Total packets per second in transaction\\
		
		Srate & Source to destination packets per second\\
		
		Drate & Destination to source packets per second \\
		
		\bottomrule
	\end{tabular}
	
\end{specialtable}

One important part of examining the dataset involves checking the representation of the classes in the {dataset,} i.e. whether one class is over or under represented, as this can have a detrimental effect on the experiments. Table \ref{DSLD} shows the amount of attack data and no attack data for each dataset used in the experiments.

\begin{specialtable}[H]
	\caption{Dataset label distribution.}\label{DSLD}
		\setlength{\tabcolsep}{7.1mm}
	\begin{tabular}{cccc}
		\toprule 
		\textbf{Dataset } & \textbf{No Attack Data} & \textbf{Attack Data }& \textbf{Total}\\
		\midrule 
		Data exfiltration & 24 & 118& 142\\
		
		DDoS HTTP & 55 & \hl{19771} & \hl{19826}\\
		
		DDoS TCP & 32 & \hl{1048543} & \hl{1048575}\\
		
		DDoS UDP &36 & \hl{1048539} &\hl{1048575}\\
		
		Key logging & 164 & 1469 & 1633\\
		
		OS Scan & 3949 & \hl{358275} &\hl{362224}\\
		
		Service scan &1993 &\hl{1046582} &\hl{1048575}\\
		
		DoS HTTP &56 &\hl{29706} &\hl{29762}\\
		
		DoS TCP & 106 &\hl{1048469} &\hl{1048575}\\
		
		DoS UDP & 37&\hl{1048538} &\hl{1048575}\\ 
		\bottomrule
	\end{tabular}
	
\end{specialtable}

   To conduct multi-class testing, a new CSV file is created using the binary classification datasets. The datasets were collected and then randomized and put into a new file. Due to the large size of the dataset, only a selected percentage of the data is used to prevent excessive run times. Table \ref{MULTIDATA} shows the class representation of the training and test data in the multi-class dataset. It is observable in both the binary and multi-class datasets  that not all   classes have equal representation. Testing with weighted classes can be done to see the effects of having equal representation among the classes. The models SVM, DT, RF, ANN,  and LR are able to use the balanced weighted classes option, which applies to the class weights as follows:

\begin{equation}
W = \frac{Samples}{Classes \times Y}
\end{equation}
where $Samples$ is the number of rows in the dataset, $Classes$ is the number of classes in the dataset, and $Y$ is the number of labels. 

\begin{specialtable}[H]
	\caption{Multi-class data representation.}\label{MULTIDATA}
		\setlength{\tabcolsep}{8.2mm}
	\begin{tabular}{cccc}
		\toprule 
		\textbf{Classes} & \textbf{Training Data} &\textbf{Test Data }& \textbf{Total}\\
		\midrule 
		No attack & 1398& 335&1733\\
		
		Data exfiltration & 22& 7& 29 \\
		
		DDoS HTTP & 4209& 1015& 5224 \\
		
		DDoS TCP & \hl{221638}& \hl{56377}& \hl{278015} \\
		
		DDoS UDP & 222728& 55302& 278030\\
		
		Key logging & 314& 81& 395\\
		
		OS Scan & 75877& 18907& 94784\\
		
		Service scan & 221745& 55768&277509\\
		
		DoS HTTP & 6343& 1475&7818\\
		
		DoS TCP & 223555& 55236& 278791\\
		
		DoS UDP & 222171& 55501& 277672\\
		\bottomrule
	\end{tabular}
	
\end{specialtable}

\subsection{Implementation}

\subsubsection{Tools Used}
We use Python version 3.7.4 programming language for the implementation of ML algorithms. The two main modules used for the implementation of the models are sklearn (also referred to as  scikit-learn) and Keras. Keras is used to implement the ANN while sklearn is used to implement the other models. {It is to be noted that, for comparison purposes, we   used the default values of hyperparameters for each   classifier.} Table \ref{MUD} contains names of the modules used and a brief description of the module.

\begin{specialtable} [H]
	\caption{Modules used    and    description.}\label{MUD}
	\setlength{\tabcolsep}{5.1mm}
	\begin{tabular}{cc}
		\toprule 
		\textbf{Module Name} & \textbf{Description}\\
		\midrule 
		numpy & Used to store the dataset in an array\\
		pandas & Used to read the dataset CSV file\\
		preprocessing & Used to normalize feature data\\
		model\_selection & Used for splitting the training and test data\\
		random & Used to randomize the multi-class dataset\\
		metrics & Contains the performance metrics used in testing the model\\
		neighbors & Contains KNN model\\
			SVM \cite{noble2006support} & Contains the SVM model\\
		tree & Contains the DT model\\
		naive bayes & Contains the NB model\\
		ensemble & Contains the RF model\\
		linear model & Contains the LR model\\
		models & Contains the ANN model\\
		layers & Contains ANN layers\\
		utils & Contains class weight for ANN\\
		\bottomrule
	\end{tabular}
	
\end{specialtable}

\subsubsection{Feature Extraction}
The dataset contains features that either contain no information or have information that is irrelevant in helping the model classify the data. The unwanted features can be removed during the preprocessing stage using the pandas module. Several features, such as flgs, proto, dir, state, saddr, daddr, srcid, smac, dmac, soui, doui, sco, record, category, and subcategory, were removed from the dataset.

\subsubsection{Feature Scaling}
The features in the dataset contain large numbers that vary in size. Therefore, it is important to normalize the data in the features. This is done by re-scaling the values of the features to within a defined scale such as $-$1 to 1 and can be defined as follows:

\begin{equation}
x' = a + \frac{(x - min(x)) (b -a)}{max(x) - min(x)}
\end{equation}
where $x'$ is the normalized value, $x$ is the original value, and $a$ and $b$ are the minimum and maximum values. The result of this will take any number between $-$1 and 1. This can be done in  Python using the MinMaxScaler in the preproccesing module.

\subsubsection{{Multi-Class Dataset}}

The multi-class dataset is created by collecting all the rows of all the datasets and  then       randomizing the rows  using the random  Python module. The random module contains the shuffle method, which allows an array, in this case the rows of the dataset, to be randomized. Due to the large size of the dataset when using it for testing, only roughly 25\% of the dataset is used, which is 1,500,000 rows. 
\subsubsection{Training Data}
The data used by the model to learn are  called the training data. Data can be split into training and test data with multiple ratios. For this study, a split of 80:20  was  used, with 80\% being used for training the models, which  is governed by the Pareto principle  that states that 80\% of result comes from 20\% of the effort. 
\subsubsection{Test Data}
Twenty percent  of the data is used for testing, which is typically a good amount of data. However, if the dataset is small,  this can result in a low amount of test data and  in the illusion that the model has done extremely well when in fact it has not had enough data to be properly tested.    To split the dataset into training and test data, train\_test\_split can be used from the  Python module named model\_selection. When using this function, the random state parameter can be used that sets the seed of the pseudo random number generator; in this case, the number 121  was  used.

\subsection{Results and Discussion}
   To test several ML algorithms and to identify which are the best and worst for classifying attack data on IoT networks, this section provides all the results and analysis based on several performance metrics including binary and multi-class testing.

\subsubsection{Binary Classification}
\textbf{\hl{Data Exfiltration:}}
Table \ref{DXR} shows the results for data exfiltration data where RF has the best scores for all the performance metrics including log loss. Whereas DT also has perfect scores,  it has a high log loss, indicating that the RF model is more confident in making predictions.
\begin{specialtable}[H]
	\caption{Data exfiltration results.}\label{DXR}
		\setlength{\tabcolsep}{1.95mm}
\begin{tabular}{ccccccc}
		\toprule 
		\textbf{Algorithms Used} & \textbf{Accuracy} & \textbf{Precision} & \textbf{Recall }& \textbf{F1 Score} & \textbf{Log Loss} & \textbf{ROC AUC}\\
		\midrule 
	KNN \cite{harrison2019} & 0.86 & 0.95 & 0.87 & 0.91 & 0.19 & 0.83\\
		
		SVM \cite{noble2006support} & 0.89 & 0.95 & 0.91 & 0.93 & 0.27 & 0.85\\
		
		DT \cite{sharmakumar2016} & 1.0 & 1.0 & 1.0 & 1.0 & 9.99 & 1.0\\ 
		
		NB \cite{rish2001empirical} & 0.89 & 1.0 & 0.87 & 0.93 & 3.57 & 0.93\\
		
		RF \cite{farnaazjabbar2016} & 1.0 & 1.0 & 1.0 & 1.0 & 0.059 & 1.0\\
		
		ANN \cite{saritas2019performance} & 0.82 & 0.82 & 1.0 & 0.90 & 2.57 & 0.5\\
		
		LR \cite{7060117} & 0.89 & 0.95 & 0.91 & 0.93 & 0.22 & 0.85\\
		\bottomrule 
	\end{tabular}
\end{specialtable}

Table \ref{DXDT} shows the confusion matrix for RF and shows two noteworthy pieces of information. The first is that the amount of data tested is very low and that the classes do not have equal representation. It is possible that the low amount of test data is having an impact on the results. However, the other models except from DT have relatively poor scores compared to RF.

\begin{specialtable}[H]
	\caption{Data exfiltration RF confusion matrix.}\label{DXDT}
		\setlength{\tabcolsep}{14.1mm}
	\begin{tabular}{ccc}
			\toprule 
		& \textbf{Actual Label }&\\
	\toprule 
		Predicted label & No Attack & Attack\\
	\midrule 
		No Attack & 5 & 0\\
		Attack & 0 & 24\\
	\bottomrule 
	\end{tabular}
	
\end{specialtable}

Table \ref{DXDTTDA} shows that increasing the test data to 30\% has a decrease in the log loss, indicating that the model performs better with more data although only marginally. Once the test data reaches 40\% and beyond, the results begin to get worse, although the model is able to maintain perfect recall with up to a 50\% split in the training and test data.
\begin{specialtable}[H]
	\caption{Data exfiltration RF test data amounts.}\label{DXDTTDA}
	\setlength{\tabcolsep}{2.42mm}
	\begin{tabular}{ccccccc}
		\toprule 
		\textbf{Test Amount} & \textbf{Accuracy }& \textbf{Precision} & \textbf{Recall }& \textbf{F1 Score} & \textbf{Log Loss} & \textbf{ROC AUC}\\
		\midrule 
		20 & 1.0 & 1.0 & 1.0 & 1.0 & 0.059 & 1.0\\ 
		
		30 & 1.0 & 1.0 & 1.0 & 1.0 & 0.043 & 1.0\\ 
		
		40 & 0.98 & 0.97 & 1.0 & 0.98 & 0.042 & 0.94\\ 
		
		50 & 0.97 & 0.96 & 1.0 & 0.98 & 0.083 & 0.9\\ 
		
		60 & 0.94 & 0.97 & 0.95 & 0.96 & 0.089 & 0.89\\
		\bottomrule 
	\end{tabular}
	
\end{specialtable}

Due to the class representation being imbalanced, the weighted classes parameter can be used. This allows the disparity of the classes to be rectified, the results of which are shown in Table \ref{DXWCL}. This option is not available when using the KNN and NB models. It is observable in Table \ref{DXWCL} that SVM has had its performance increase by using weighted classes with all metrics increasing and log loss decreasing. ANN is unaffected by weighted classes and LR is marginally affected with the model perfect precision but lowering its recall. DT losses its perfect scores while RF is able to keep perfect scores but slightly increases its log loss.

\begin{specialtable}[H]
	\caption{Data exfiltration weighted classes results.}\label{DXWCL}
		\setlength{\tabcolsep}{1.95mm}
\begin{tabular}{ccccccc}
		\toprule 
		\textbf{Algorithms Used} & \textbf{Accuracy} & \textbf{Precision} & \textbf{Recall} & \textbf{F1 Score }& \textbf{Log Loss} & \textbf{ROC AUC}\\
		\midrule 
		KNN \cite{harrison2019} & n/a & n/a & n/a & n/a & n/a & n/a\\
		
			SVM \cite{noble2006support} & 0.93 & 1.0 & 0.91 & 0.95 & 0.25 & 0.95\\
		
		DT \cite{sharmakumar2016} & 0.93 & 1.0 & 0.91 & 0.95 & 0.12 & 0.95\\ 
		
		NB \cite{rish2001empirical} & n/a & n/a & n/a & n/a & n/a & n/a\\
		
		RF \cite{farnaazjabbar2016} & 1.0 & 1.0 & 1.0 & 1.0 & 0.074 & 1.0\\
		
		ANN \cite{saritas2019performance} & 0.82 & 0.82 & 1.0 & 0.90 & 2.57 & 0.5\\
		
		LR \cite{7060117} & 0.89 & 1.0 & 0.87 & 0.93 & 0.38 & 0.93\\
		\bottomrule 
	\end{tabular}
	
\end{specialtable}

Without using weighted classes, RF is the best model due to its low log loss when compared to DT. When weighted classes are applied, RF is still the best model with perfect scores and a low log loss, indicating that the model is confident in making predictions.

\textbf{DDoS HTTP:}
Table \ref{DDHR} shows the results of DDoS HTTP data. DT has perfect performance scores but a high log loss of 7.25. This dataset does not suffer from a lack of data, rather it suffers from a large imbalance of data since the attack data have more prevalence in the dataset, as shown in Table \ref{DHDCM}.
\begin{specialtable}[H]
	\caption{DDoS HTTP results.}\label{DDHR}
		\setlength{\tabcolsep}{1.95mm}
\begin{tabular}{ccccccc}
		\toprule 
		\textbf{Algorithms Used} & \textbf{Accuracy} & \textbf{Precision} & \textbf{Recall }& \textbf{F1 Score} & \textbf{Log Loss} & \textbf{ROC AUC}\\
		\midrule 
		KNN \cite{harrison2019} & 0.99 & 0.99 & 1.0 & 0.99 & 0.0095 & 0.83\\
		
			SVM \cite{noble2006support} & 0.99 & 0.99 & 1.0 & 0.99 & 0.0093 & 0.77\\
		
		DT \cite{sharmakumar2016} & 1.0 & 1.0 & 1.0 & 1.0 & 7.25 & 1.0\\ 
		
		NB \cite{rish2001empirical} & 0.99 & 0.99 & 0.99 & 0.99 & 0.063 & 0.66\\
		
		RF \cite{farnaazjabbar2016} & 0.99 & 0.99 & 1.0 & 0.99 & 0.0021 & 0.88\\
		
		ANN \cite{saritas2019performance} & 0.99 & 0.99 & 1.0 & 0.99 & 0.044 & 0.5\\
		
		LR \cite{7060117} & 0.99 & 0.99 & 1.0 & 0.99 & 0.0069 & 0.77\\
		\bottomrule 
	\end{tabular}
	
\end{specialtable}

\begin{specialtable}[H]
	\caption{DDoS HTTP DT confusion matrix.}\label{DHDCM}
		\setlength{\tabcolsep}{14.1mm}
	\begin{tabular}{ccc}
		\toprule 
		& \textbf{Actual Label }&\\
\midrule 
		Predicted label & No Attack & Attack\\
	\midrule 
		No Attack & 9 & 0\\
		Attack & 0 & 3950\\
		\bottomrule 
	\end{tabular}
	
\end{specialtable}

This confusion matrix shows a large disparity in the data with a ratio of 3:1319 in favor of attack data. A large disparity in the dataset can cause the log loss to be affected, as log loss is based on probability, and, because the data  are  more likely to be attack data, this can result in a skewed log loss. 

Table \ref{DDHWCR} shows the results of weighted classes on the DDoS HTTP data. With weighted classes, both SVM and LR have a sizeable  decrease in performance across all metrics except log loss which has decreased for both and ROC AUC, which has increased for both. ANN is unaffected by the weighted classes and retains its perfect recall, whereas RF loses the perfect recall. DT loses its perfect scores but has a large decrease in its log loss.

\begin{specialtable} [H]
	\caption{DDoS HTTP weighted classes results.}\label{DDHWCR}
	\setlength{\tabcolsep}{1.95mm}
	\begin{tabular}{ccccccc}
		\toprule 
		\textbf{Algorithms Used} & \textbf{Accuracy }& \textbf{Precision} & \textbf{Recall} & \textbf{F1 Score} & \textbf{Log Loss} & \textbf{ROC AUC}\\
		\midrule 
		KNN \cite{harrison2019} & n/a & n/a & n/a & n/a & n/a & n/a\\
		
			SVM \cite{noble2006support} & 0.89 & 0.99 & 0.89 & 0.94 & 0.013 & 0.83\\
		
		DT \cite{sharmakumar2016} & 0.99 & 0.99 & 0.99 & 0.99 & 0.018 & 0.88\\ 
		
		NB \cite{rish2001empirical} & n/a & n/a & n/a & n/a & n/a & n/a\\
		
		RF \cite{farnaazjabbar2016} & 0.99 & 0.99 & 0.99 & 0.99 & 0.0047 & 0.88\\
		
		ANN \cite{saritas2019performance} & 0.99 & 0.99 & 1.0 & 0.99 & 0.044 & 0.5\\
		
		LR \cite{7060117} & 0.91 & 0.99 & 0.91 & 0.95 & 0.15 & 0.90\\
		\bottomrule 
	\end{tabular}
	
\end{specialtable}	

Without using weighted classes, DT is the best model due to the perfect scores, although the high log loss is a factor to consider. RF would be the second best as it has perfect recall as well as the lowest log loss and the highest ROC AUC. When weighted classes are applied, ANN is the best model as it has perfect recall and a low log loss.

\textbf{\hl{DDoS TCP:}}
Table \ref{DDTR} shows the results of the DDoS TCP data. The models DT and RF both have perfect score except for log loss which is high for both. Table \ref{DDTCON} shows the confusion matrix for RF and once again the matrix shows a very large disparity in the data represented.
\begin{specialtable} [H]
	\caption{DDoS TCP results.}\label{DDTR}
	\setlength{\tabcolsep}{1.95mm}
	\begin{tabular}{ccccccc}
		\toprule 
		\textbf{Algorithms Used} & \textbf{Accuracy }& \textbf{Precision} & \textbf{Recall} & \textbf{F1 Score} & \textbf{Log Loss} & \textbf{ROC AUC}\\
		\midrule 
		KNN \cite{harrison2019} & 0.99 & 0.99 & 1.0 & 0.99 & 1.76 & 0.83\\
		
			SVM \cite{noble2006support} & 0.99 & 1.0 & 0.99 & 0.99 & 5.82 & 0.83\\
		
		DT \cite{sharmakumar2016} & 1.0 & 1.0 & 1.0 & 1.0 & 9.99 & 1.0\\ 
		
		NB \cite{rish2001empirical} & 0.99 & 1.0 & 0.99 & 0.99 & 0.029 & 0.99\\
		
		RF \cite{farnaazjabbar2016} & 1.0 & 1.0 & 1.0 & 1.0 & 2.55 & 1.0\\
		
		ANN \cite{saritas2019performance} & 0.99 & 0.99 & 1.0 & 0.99 & 4.75 & 0.5\\
		
		LR \cite{7060117} & 0.99 & 0.99 & 1.0 & 0.99 & 0.00010 & 0.58\\
		\bottomrule 
	\end{tabular}
	
\end{specialtable}

\begin{specialtable}[H]
	\caption{DDoS TCP RF confusion matrix.}\label{DDTCON}
	\setlength{\tabcolsep}{14.1mm}
	\begin{tabular}{ccc}
			\toprule
		& \textbf{Actual Label} &\\
			\midrule 
		Predicted label & No Attack & Attack\\
			\midrule 
		No Attack & 6 & 0\\
		Attack & 0 & \hl{209709}\\
		\bottomrule 
	\end{tabular}
	
\end{specialtable}

Table \ref{DDTWCR} shows the results of DDoS TCP data with weighted classes enabled. With weighted classes enabled, SVM has lost its perfect precision but lowered its log loss significantly. DT and ANN are unaffected by the weighted classes but RF retains its perfect scores and lowers its log loss slightly. LR has lost its perfect recall and increased its log loss and ROC AUC.

\begin{specialtable}[H]
	\caption{DDoS TCP weighted classes results.}\label{DDTWCR}
		\setlength{\tabcolsep}{1.95mm}
	\begin{tabular}{ccccccc}
		\toprule 
		\textbf{Algorithms Used} & \textbf{Accuracy }& \textbf{Precision} & \textbf{Recall} & \textbf{F1 Score} & \textbf{Log Loss} & \textbf{ROC AUC}\\
		\midrule 
		KNN \cite{harrison2019} & n/a & n/a & n/a & n/a & n/a & n/a\\
		
			SVM \cite{noble2006support} & 0.99 & 0.99 & 0.99 & 0.99 & 0.00040 & 0.83\\
		
		DT \cite{sharmakumar2016} & 1.0 & 1.0 & 1.0 & 1.0 & 9.99 & 1.0\\ 
		
		NB \cite{rish2001empirical} & n/a & n/a & n/a & n/a & n/a & n/a\\
		
		RF \cite{farnaazjabbar2016} & 1.0 & 1.0 & 1.0 & 1.0 & 1.33 & 1.0\\
		
		ANN \cite{saritas2019performance} & 0.99 & 0.99 & 1.0 & 0.99 & 4.75 & 0.5\\
		
		LR \cite{7060117} & 0.99 & 0.99 & 0.99 & 0.99 & 0.025 & 0.91\\
		\bottomrule 
	\end{tabular}
	
\end{specialtable}

Both with and without weighted classes, RF is the best model as it has perfect scores. With weighed classes, the log loss is lowered but is still quite high when compared to LR which has a very low log loss.

\textbf{\hl{DDoS UDP:}}
Table \ref{DDUR} shows the results of the DDoS UDP data, where both KNN and and DT have perfect score but KNN is the better model as it has a lower log loss. Although the log loss is still high, this is the case for all the models apart from NB. Table \ref{DDUCON} shows the confusion matrix for RF, which shows the disparity in the class representation.
\begin{specialtable}[H]
	\caption{DDoS UDP results.}\label{DDUR}
		\setlength{\tabcolsep}{1.95mm}
	\begin{tabular}{ccccccc}
		\toprule 
		\textbf{Algorithms Used} & \textbf{Accuracy }& \textbf{Precision} & \textbf{Recall} & \textbf{F1 Score} & \textbf{Log Loss} & \textbf{ROC AUC}\\
		\midrule 
		KNN \cite{harrison2019} & 1.0 & 1.0 & 1.0 & 1.0 & 4.56 & 1.0\\
		
			SVM \cite{noble2006support} & 0.99 & 0.99 & 1.0 & 0.99 & 8.93 & 0.92\\
		
		DT \cite{sharmakumar2016} & 1.0 & 1.0 & 1.0 & 1.0 & 9.99 & 1.0\\ 
		
		NB \cite{rish2001empirical} & 0.99 & 1.0 & 0.99 & 0.99 & 0.00098 & 0.99\\
		
		RF \cite{farnaazjabbar2016} & 0.99 & 0.99 & 1.0 & 0.99 & 5.71 & 0.92\\
		
		ANN \cite{saritas2019performance} & 0.99 & 0.99 & 1.0 & 0.99 & 5.30 & 0.5\\
		
		LR \cite{7060117} & 0.99 & 0.99 & 1.0 & 0.99 & 7.77 & 0.78\\
		\bottomrule 
	\end{tabular}
	
\end{specialtable}

\begin{specialtable}[H]
	\caption{DDoS UDP KNN confusion matrix.}\label{DDUCON}
		\setlength{\tabcolsep}{14.1mm}
	\begin{tabular}{ccc}
	\toprule
& \textbf{Actual Label} &\\
\midrule
		Predicted label & No Attack & Attack\\
			\midrule 
		No Attack & 7 & 0\\
		Attack & 0 & \hl{209708}\\
		\bottomrule
	\end{tabular}
	
\end{specialtable}

Table \ref{DDUWCR} shows the results of DDoS UDP data with weighted classes enabled. The table shows that SVM has gained perfect scores and lowered it loss loss, while DT has lost its perfect scores and lowered its log loss substantially. RF has gained perfect scores and lowered its log loss, while ANN is unaffected. LR has lost perfect recall but gained perfect precision and lowered its log loss and increased its ROC AUC.

\begin{specialtable}[H]
	\caption{DDoS UDP weighted classes results.}\label{DDUWCR}
		\setlength{\tabcolsep}{1.95mm}
	\begin{tabular}{ccccccc}
		\toprule 
		\textbf{Algorithms Used} & \textbf{Accuracy }& \textbf{Precision} & \textbf{Recall} & \textbf{F1 Score} & \textbf{Log Loss} & \textbf{ROC AUC}\\
		\midrule 
		KNN \cite{harrison2019} & n/a & n/a & n/a & n/a & n/a & n/a\\
		
			SVM \cite{noble2006support} & 1.0 & 1.0 & 1.0 & 1.0 & 2.84 & 1.0\\
		
		DT \cite{sharmakumar2016} & 0.99 & 1.0 & 0.99 & 0.99 & 0.000011 & 0.99\\ 
		
		NB \cite{rish2001empirical} & n/a & n/a & n/a & n/a & n/a & n/a\\
		
		RF \cite{farnaazjabbar2016} & 1.0 & 1.0 & 1.0 & 1.0 & 0.0020 & 1.0\\
		
		ANN \cite{saritas2019performance} & 0.99 & 0.99 & 1.0 & 0.99 & 5.30 & 0.5\\
		
		LR \cite{7060117} & 0.99 & 1.0 & 0.99 & 0.99 & 0.00028 & 0.99\\
		\bottomrule 
	\end{tabular}
	
\end{specialtable}

Without weighted classes, KNN is the best model as it has perfect scores but the log loss is high. NB would be second best as it has perfect precision and a low log loss. With weighted classes, RF is the best model as it has perfect scores and a low log loss. 

\textbf{\hl{Key logging:}}
Table \ref{KLR} shows the results of Key logging data. DT is the best model as it has the best log loss and ROC AUC scores combined with perfect precision while having high metric scores. 

\begin{specialtable}[H]
	\caption{Key logging results.}\label{KLR}
	\setlength{\tabcolsep}{1.95mm}
	\begin{tabular}{ccccccc}
		\toprule 
		\textbf{Algorithms Used} & \textbf{Accuracy }& \textbf{Precision} & \textbf{Recall} & \textbf{F1 Score} & \textbf{Log Loss} & \textbf{ROC AUC}\\
		\midrule 
		KNN \cite{harrison2019} & 0.98 & 0.98 & 1.0 & 0.99 & 0.33 & 0.93\\
		
			SVM \cite{noble2006support} & 0.96 & 0.96 & 1.0 & 0.98 & 0.16 & 0.81\\
		
		DT \cite{sharmakumar2016} & 0.99 & 1.0 & 0.99 & 0.99 & 0.0085 & 0.99\\ 
		
		NB \cite{rish2001empirical} & 0.91 & 0.92 & 0.98 & 0.95 & 2.64 & 0.58\\
		
		RF \cite{farnaazjabbar2016} & 0.99 & 0.99 & 1.0 & 0.99 & 0.022 & 0.96\\
		
		ANN \cite{saritas2019performance} & 0.91 & 0.91 & 1.0 & 0.95 & 1.58 & 0.5\\
		
		LR \cite{7060117} & 0.96 & 0.96 & 1.0 & 0.98 & 0.16 & 0.79\\
		\bottomrule 
	\end{tabular}
\end{specialtable}
Table \ref{KLCON} shows the confusion matrix for DT where it is observable that the dataset has a low amount of data and the data are imbalanced.

\begin{specialtable}[H]
	\caption{Key logging DT confusion matrix.}\label{KLCON}
		\setlength{\tabcolsep}{14.1mm}
	\begin{tabular}{ccc}
				\toprule 
		& \textbf{Actual Label} &\\
			\midrule 
		Predicted label & No Attack & Attack\\
			\midrule 
		No Attack & 29 & 0\\
		Attack & 2 & 296\\
		\bottomrule
	\end{tabular}
	
\end{specialtable}

Just as with data exfiltration, the amount of test data can be increased to observe the effect on the scores of the DT model. Table \ref{KLRFTDA} shows the results of increasing the test data for key logging data. Increasing the test data to 30\% gives the model perfect recall instead of perfect accuracy. Once the data are increased to 50\%, the model no longer has perfect recall or precision. Based on the changes in the results, it is observable that the low amount of data   has a significant impact on the results of the model.

\begin{specialtable}[H]
	\caption{Key logging DT test data amounts.}\label{KLRFTDA}
		\setlength{\tabcolsep}{2.42mm}
	\begin{tabular}{ccccccc}
		\toprule 
		\textbf{Test Amount }& \textbf{Accuracy }& \textbf{Precision} & \textbf{Recall }& \textbf{F1 Score} & \textbf{Log Loss }& \textbf{ROC AUC}\\
		\midrule 
		20 & 0.99 & 1.0 & 0.99 & 0.99 & 0.0085 & 0.99\\ 
		
		30 & 0.99 & 0.99 & 1.0 & 0.99 & 0.080 & 0.96\\ 
		
		40 & 0.99 & 0.98 & 1.0 & 0.99 & 0.11 & 0.95\\ 
		
		50 & 0.99 & 0.99 & 0.99 & 0.99 & 0.13 & 0.95\\ 
		
		\bottomrule 
	\end{tabular}
	
\end{specialtable}

Table \ref{KLWCR} shows the results of key logging data with weighted classes enabled. SVM shows an overall decrease in performance with the model no longer having perfect recall. DT and RF also show a drop in performance with the models losing their perfect precision and recall, respectively. ANN is unaffected with LR having a large decrease in the models recall leading to the worst performance of all the models.

\begin{specialtable}[H]
	\caption{Key logging weighted classes results.}\label{KLWCR}
		\setlength{\tabcolsep}{1.95mm}
	\begin{tabular}{ccccccc}
		\toprule 
		\textbf{Algorithms Used} & \textbf{Accuracy }& \textbf{Precision} & \textbf{Recall} & \textbf{F1 Score} & \textbf{Log Loss} & \textbf{ROC AUC}\\
		\midrule 
		KNN \cite{harrison2019} & n/a & n/a & n/a & n/a & n/a & n/a\\
		
			SVM \cite{noble2006support} & 0.87 & 0.98 & 0.87 & 0.92 & 0.17 & 0.88\\
		
		DT \cite{sharmakumar2016} & 0.98 & 0.99 & 0.98 & 0.99 & 0.038 & 0.97\\ 
		
		NB \cite{rish2001empirical} & n/a & n/a & n/a & n/a & n/a & n/a\\
		
		RF \cite{farnaazjabbar2016} & 0.98 & 0.99 & 0.98 & 0.99 & 0.051 & 0.97\\
		
		ANN \cite{saritas2019performance} & 0.91 & 0.91 & 1.0 & 0.95 & 1.58 & 0.5\\
		
		LR \cite{7060117} & 0.77 & 0.98 & 0.76 & 0.85 & 0.46 & 0.82\\
		\bottomrule 
	\end{tabular}
	
\end{specialtable}

Without weighted classes, DT is the best model with the lowest log loss and highest ROC AUC as well as perfect precision. With weighted class, all the models tested had a decrease in performance except for ANN, which was unchanged. Apart from the models perfect recall, it still has comparatively worse scores than DT and RF. Unless perfect recall is a factor DT should be used as it will correctly classify more data than ANN. 

\textbf{\hl{OS Scan:}}
Table \ref{OSSR} shows the results for OS Scan data. All of the models have good scores with RF, ANN and LR having a perfect recall indicating the models made no false negatives. RF has a higher precision than LR and ANN as well as having a lower log loss and higher ROC AUC. This would suggest that RF is the best model. However, inspection of the confusion matrix shows a large imbalance of data in the dataset, as shown in Table \ref{osscanrf}.

\begin{specialtable}[H]
	\caption{OS Scan results.}\label{OSSR}
		\setlength{\tabcolsep}{1.95mm}
	\begin{tabular}{ccccccc}
		\toprule 
		\textbf{Algorithms Used} & \textbf{Accuracy }& \textbf{Precision} & \textbf{Recall} & \textbf{F1 Score} & \textbf{Log Loss} & \textbf{ROC AUC}\\
		\midrule 
		KNN \cite{harrison2019} & 0.99 & 0.99 & 0.99 & 0.99 & 0.063 & 0.80\\
		
			SVM \cite{noble2006support} & 0.94 & 0.99 & 0.94 & 0.97 & 0.024 & 0.96\\
		
		DT \cite{sharmakumar2016} & 0.99 & 0.99 & 0.99 & 0.99 & 0.0038 & 0.98\\ 
		
		NB \cite{rish2001empirical} & 0.98 & 0.98 & 0.99 & 0.99 & 0.54 & 0.51\\
		
		RF \cite{farnaazjabbar2016} & 0.99 & 0.99 & 1.0 & 0.99 & 0.0061 & 0.83\\
		
		ANN \cite{saritas2019performance} & 0.98 & 0.98 & 1.0 & 0.99 & 0.16 & 0.5\\
		
		LR \cite{7060117} & 0.98 & 0.98 & 1.0 & 0.99 & 0.036 & 0.50\\
		\bottomrule 
	\end{tabular}
	
\end{specialtable}

\begin{specialtable}[H]
	\caption{OS Scan RF confusion matrix.}\label{osscanrf}
		\setlength{\tabcolsep}{14.1mm}
	\begin{tabular}{ccc}
		\toprule 
		& \textbf{Actual Label} &\\
			\midrule 
		Predicted label & No Attack & Attack\\
			\midrule 
		No Attack & 608 & 161\\
		Attack & 0 & \hl{71673}\\
				\bottomrule 
	\end{tabular}
\end{specialtable}

Table \ref{OSSWCR} shows the results of OS scan data with weighted classes enabled. SVM shows a decrease in accuracy, recall, F1 score, log loss, and ROC AUC. The table also shows that the models decreased the performance overall. DT shows a decrease in log loss and ROC AUC marking a slight increase in the models confidence but lower ability to perform well at different thresholds. RF has lost its perfect recall and has an increased log loss and ROC AUC. ANN has seen no change to its results, whereas LR has a large performance decrease with only ROC AUC have been improved.

\begin{specialtable}[H]
	\caption{OS scan weighted classes results.}\label{OSSWCR}
		\setlength{\tabcolsep}{1.95mm}
	\begin{tabular}{ccccccc}
		\toprule 
		\textbf{Algorithms Used} & \textbf{Accuracy }& \textbf{Precision} & \textbf{Recall} & \textbf{F1 Score} & \textbf{Log Loss} & \textbf{ROC AUC}\\
		\midrule 
		KNN \cite{harrison2019} & n/a & n/a & n/a & n/a & n/a & n/a\\
		
			SVM \cite{noble2006support} & 0.89 & 0.99 & 0.89 & 0.94 & 0.013 & 0.83\\
		
		DT \cite{sharmakumar2016} & 0.99 & 0.99 & 0.99 & 0.99 & 0.025 & 0.88\\ 
		
		NB \cite{rish2001empirical} & n/a & n/a & n/a & n/a & n/a & n/a\\
		
		RF \cite{farnaazjabbar2016} & 0.99 & 0.99 & 0.99 & 0.99 & 0.030 & 0.99\\
		
		ANN \cite{saritas2019performance} & 0.98 & 0.98 & 1.0 & 0.99 & 0.16 & 0.5\\
		
		LR \cite{7060117} & 0.90 & 0.99 & 0.90 & 0.95 & 0.19 & 0.94\\
		\bottomrule 
	\end{tabular}
	
\end{specialtable}

Without weighted classes, RF is the best model as it has perfect recall and the lowest log loss also having the highest ROC AUC. With weighted classes, ANN is the only model with perfect recall but DT and RF both have better accuracy, precision, log loss, and ROC AUC. If having no false positives is needed, then ANN is the best, but DT is better at classifying data in general. 

\textbf{\hl{Service Scan:}}
Table \ref{SSR} shows the results for service scan data. The models SVM, RF, and ANN have perfect recall but have poor ROC AUC scores. DT has the highest ROC AUC and the lowest log loss, but RF could be considered the best due to its perfect recall.
\begin{specialtable}[H]
	\caption{Service Scan results.}\label{SSR}
		\setlength{\tabcolsep}{1.95mm}
	\begin{tabular}{ccccccc}
		\toprule 
		\textbf{Algorithms Used} & \textbf{Accuracy }& \textbf{Precision} & \textbf{Recall} & \textbf{F1 Score} & \textbf{Log Loss} & \textbf{ROC AUC}\\
		\midrule 
		KNN \cite{harrison2019} & 0.99 & 0.99 & 0.99 & 0.99 & 0.013 & 0.79\\
		
			SVM \cite{noble2006support} & 0.99 & 0.99 & 1.0 & 0.99 & 0.012 & 0.54\\
		
		DT \cite{sharmakumar2016} & 0.99 & 0.99 & 0.99 & 0.99 & 0.0028 & 0.84\\ 
		
		NB \cite{rish2001empirical} & 0.99 & 0.99 & 0.99 & 0.99 & 0.26 & 0.58\\
		
		RF \cite{farnaazjabbar2016} & 0.99 & 0.99 & 1.0 & 0.99 & 0.0039 & 0.54\\
		
		ANN \cite{saritas2019performance} & 0.99 & 0.99 & 1.0 & 0.99 & 0.029 & 0.5\\
		
		LR \cite{7060117} & 0.99 & 0.99 & 0.99 & 0.99 & 0.0087 & 0.54\\
		\bottomrule 
	\end{tabular}
	
\end{specialtable}

Table \ref{SSRFCON} shows the confusion matrix for RF  as well as   the imbalanced data.

\begin{specialtable}[H]
	\caption{Service Scan RF confusion matrix.}\label{SSRFCON}
		\setlength{\tabcolsep}{14.1mm}
	\begin{tabular}{ccc}
				\toprule
		& \textbf{Actual Label} &\\ 
			\midrule
		Predicted label & No attack & Attack\\
		\midrule
		No attack & 31 & 350\\
		Attack & 0 & \hl{209334}\\ 
			\bottomrule 
	\end{tabular}
\end{specialtable}

Table \ref{SSWCR} shows the results of service scan data with weighted classes enabled. SVM was not tested due to excessive running times. DT, RF, and LR have increased their ROC AUC but all other metrics have been negatively affected. ANN is unaffected, being the only model to keep its perfect recall.

\begin{specialtable}[H]
	\caption{Service scan weighted classes results.}\label{SSWCR}
		\setlength{\tabcolsep}{1.95mm}
	\begin{tabular}{ccccccc}
		\toprule 
		\textbf{Algorithms Used} & \textbf{Accuracy }& \textbf{Precision} & \textbf{Recall} & \textbf{F1 Score} & \textbf{Log Loss} & \textbf{ROC AUC}\\
		\midrule 
		KNN \cite{harrison2019} & n/a & n/a & n/a & n/a & n/a & n/a\\
		
			SVM \cite{noble2006support} & n/a & n/a & n/a & n/a & n/a & n/a\\
		
		DT \cite{sharmakumar2016} & 0.97 & 0.99 & 0.97 & 0.98 & 0.079 & 0.97\\ 
		
		NB \cite{rish2001empirical} & n/a & n/a & n/a & n/a & n/a & n/a\\
		
		RF \cite{farnaazjabbar2016} & 0.94 & 0.99 & 0.94 & 0.97 & 0.13 & 0.96\\
		
		ANN \cite{saritas2019performance} & 0.99 & 0.99 & 1.0 & 0.99 & 0.029 & 0.5\\
		
		LR \cite{7060117} & 0.85 & 0.99 & 0.85 & 0.92 & 0.29 & 0.90\\
		\bottomrule 
	\end{tabular}
	
\end{specialtable}

Without weighted classes of the models with perfect recall, RF is the best as it has the lowest log loss and highest ROC AUC. However, \hl{DT   has} the best log loss but does not have perfect recall. With weighted classes, ANN is the best as it is the only model to retain perfect recall, but its ROC AUC is the poorest of all the models. 

\textbf{\hl{DoS HTTP:}}
Table \ref{DHR} shows the results for DoS data; DT and RF both have perfect scores and a low log loss with DT narrowly beating RF.
\begin{specialtable}[H]
	\caption{DoS HTTP results.}\label{DHR}
		\setlength{\tabcolsep}{1.95mm}
	\begin{tabular}{ccccccc}
		\toprule 
		\textbf{Algorithms Used} & \textbf{Accuracy }& \textbf{Precision} & \textbf{Recall} & \textbf{F1 Score} & \textbf{Log Loss} & \textbf{ROC AUC}\\
		\midrule 
		KNN \cite{harrison2019} & 0.99 & 0.99 & 1.0 & 0.99 & 0.0063 & 0.90\\
		
			SVM \cite{noble2006support} & 0.99 & 0.99 & 1.0 & 0.99 & 0.0065 & 0.86\\
		
		DT \cite{sharmakumar2016} & 1.0 & 1.0 & 1.0 & 1.0 & 0.00013 & 1.0\\ 
		
		NB \cite{rish2001empirical} & 0.99 & 0.99 & 0.99 & 0.99 & 0.034 & 0.77\\
		
		RF \cite{farnaazjabbar2016} & 1.0 & 1.0 & 1.0 & 1.0 & 0.00094 & 1.0\\
		
		ANN \cite{saritas2019performance} & 0.99 & 0.99 & 1.0 & 0.99 & 0.029 & 0.5\\
		
		LR \cite{7060117} & 0.99 & 0.99 & 1.0 & 0.99 & 0.0044 & 0.81\\
		\bottomrule 
	\end{tabular}
	
\end{specialtable}

Table \ref{SHRCON} shows the confusion matrix for RF which showcases the disparity in the dataset.

\begin{specialtable}[H]
	\caption{DoS HTTP RF confusion matrix.}\label{SHRCON}
		\setlength{\tabcolsep}{14.1mm}
	\begin{tabular}{ccc}
			\toprule 
		& \textbf{Actual Label} &\\ 
				\midrule 
		Predicted label & No attack & Attack\\
				\midrule 
		No attack & 11 & 0\\
		Attack & 0 & 5942\\ 
				\bottomrule 
	\end{tabular}
\end{specialtable}

Table \ref{DHWCR} shows the results of DoS HTTP data with weighted classes enabled. SVM shows deceased performance in all metric except for ROC AUC. DT and RF have lost their perfect scores and have an increased log loss. ANN is unaffected, whereas LR has seen a decrease in all performance metrics apart from ROC AUC, which has increased.

\begin{specialtable}[H]
	\caption{DoS HTTP weighted classes results.}\label{DHWCR}
		\setlength{\tabcolsep}{1.95mm}
	\begin{tabular}{ccccccc}
		\toprule 
		\textbf{Algorithms Used} & \textbf{Accuracy }& \textbf{Precision} & \textbf{Recall} & \textbf{F1 Score} & \textbf{Log Loss} & \textbf{ROC AUC}\\
		\midrule 
		KNN \cite{harrison2019} & n/a & n/a & n/a & n/a & n/a & n/a\\
		
			SVM \cite{noble2006support} & 0.90 & 0.99 & 0.90 & 0.95 & 0.0067 & 0.90\\
		
		DT \cite{sharmakumar2016} & 0.99 & 0.99 & 0.99 & 0.99 & 0.0098 & 0.95\\ 
		
		NB \cite{rish2001empirical} & n/a & n/a & n/a & n/a & n/a & n/a\\
		
		RF \cite{farnaazjabbar2016} & 0.99 & 0.99 & 0.99 & 0.99 & 0.0097 & 0.95\\
		
		ANN \cite{saritas2019performance} & 0.99 & 0.99 & 1.0 & 0.99 & 0.029 & 0.5\\
		
		LR \cite{7060117} & 0.88 & 0.99 & 0.88 & 0.94 & 0.21 & 0.89\\
		\bottomrule 
	\end{tabular}
	
\end{specialtable}

\begin{specialtable}[H]
	\caption{\hl{DoS TCP results.}}\label{DTR}
		\setlength{\tabcolsep}{1.95mm}
	\begin{tabular}{ccccccc}
		\toprule 
		\textbf{Algorithms Used} & \textbf{Accuracy }& \textbf{Precision} & \textbf{Recall} & \textbf{F1 Score} & \textbf{Log Loss} & \textbf{ROC AUC}\\
		\midrule 
		KNN \cite{harrison2019} & 0.99 & 0.99 & 1.0 & 0.99 & 0.00035 & 0.90\\
		
			SVM \cite{noble2006support} & 0.99 & 0.99 & 1.0 & 0.99 & 0.0011 & 0.61\\
		
		DT \cite{sharmakumar2016} & 0.99 & 0.99 & 1.0 & 0.99 & 1.62 & 0.95\\ 
		
		NB \cite{rish2001empirical} & 0.99 & 0.99 & 0.99 & 0.99 & 0.026 & 0.69\\
		
		RF \cite{farnaazjabbar2016} & 0.99 & 0.99 & 1.0 & 0.99 & 2.25 & 0.92\\
		
		ANN \cite{saritas2019performance} & 0.99 & 0.99 & 1.0 & 0.99 & 0.0016 & 0.5\\
		
		LR \cite{7060117} & 0.99 & 0.99 & 1.0 & 0.99 & 0.00066 & 0.61\\
		\bottomrule 
	\end{tabular}
	
\end{specialtable}

\hl{Table} \ref{DTRCON} \hl{shows the confusion }matrix for DT and shows the imbalance of the data in the dataset.

\begin{specialtable}[H]
	\caption{DoS TCP DT confusion matrix.}\label{DTRCON}
		\setlength{\tabcolsep}{14.1mm}
	\begin{tabular}{ccc}
			\toprule 
		& \textbf{Actual Label} &\\ 
			\midrule 
		Predicted label & No attack & Attack\\
			\midrule 
		No attack & 19 & 2\\
		Attack & 0 & \hl{209694}\\ 
			\bottomrule 
	\end{tabular}
\end{specialtable}

Without weighted classes, DT is the best model as it has perfect scores and the lowest log loss. With weighted classes, ANN is the best model as it has perfect recall. In regards to the models ability to classify data, ANN comes out on top due to having perfect recall. 

\textbf{\hl{DoS TCP:}}
Table \ref{DTR} shows the results for DoS TCP data, where all the models apart from NB have perfect recall. DT and RF have the best ROC AUC scores, but both have high log losses when compared to the other models. KNN has the lowest log loss and a ROC AUC almost as good as RF.

Table \ref{DTWCR} shows the results of DoS TCP data with weighted classes enabled. SVM was not recorded due to excessively long running times. With weighted classes, both DT and RF have lost their perfect recall, but DT has gained perfect precision. Both models have also seen an improvement in log loss and ROC AUC. ANN is affected and LR has had a performance decrease in almost all metrics. 
\begin{specialtable}[H]
	\caption{DoS TCP weighted classes results.}\label{DTWCR}
		\setlength{\tabcolsep}{1.95mm}
	\begin{tabular}{ccccccc}
		\toprule 
		\textbf{Algorithms Used} & \textbf{Accuracy }& \textbf{Precision} & \textbf{Recall} & \textbf{F1 Score} & \textbf{Log Loss} & \textbf{ROC AUC}\\
		\midrule 
		KNN \cite{harrison2019} & n/a & n/a & n/a & n/a & n/a & n/a\\
		
			SVM \cite{noble2006support} & n/a & n/a & n/a & n/a & n/a & n/a\\
		
		DT \cite{sharmakumar2016} & 0.99 & 1.0 & 0.99 & 0.99 & 0.018 & 0.99\\ 
		
		NB \cite{rish2001empirical} & n/a & n/a & n/a & n/a & n/a & n/a\\
		
		RF \cite{farnaazjabbar2016} & 0.99 & 0.99 & 0.99 & 0.99 & 0.022 & 0.97\\
		
		ANN \cite{saritas2019performance} & 0.99 & 0.99 & 1.0 & 0.99 & 0.0016 & 0.5\\
		
		LR \cite{7060117} & 0.96 & 0.99 & 0.96 & 0.98 & 0.078 & 0.91\\
		\bottomrule 
	\end{tabular}
	
\end{specialtable}

Without weighted classes, KNN could be considered the best model as it has the lowest log loss and a reasonably high ROC AUC. DT and RF have a higher ROC AUC but also have a considerably higher log loss than KNN. With weighted classes, both DT and ANN could be considered the best with DT having perfect precision and ANN having perfect recall. Both models also have a low log loss, but ANN has a poorer ROC AUC score. 

\textbf{\hl{DoS UDP:}}
Table \ref{DUR} shows the results for DoS UDP data. NB is the best model with perfect precision, low log loss, and high ROC AUC, as well as having high metrics across all categories. All the other models have perfect recall but   have either a high log loss or a low ROC AUC, or both. 
\begin{specialtable}[H]
	\caption{DoS UDP results.}\label{DUR}
		\setlength{\tabcolsep}{1.95mm}
	\begin{tabular}{ccccccc}
		\toprule 
		\textbf{Algorithms Used} & \textbf{Accuracy }& \textbf{Precision} & \textbf{Recall} & \textbf{F1 Score} & \textbf{Log Loss} & \textbf{ROC AUC}\\
		\midrule 
		KNN \cite{harrison2019} & 0.99 & 0.99 & 1.0 & 0.99 & 3.28 & 0.75\\
		
			SVM \cite{noble2006support} & 0.99 & 0.99 & 1.0 & 0.99 & 0.00039 & 0.68\\
		
		DT \cite{sharmakumar2016} & 0.99 & 0.99 & 1.0 & 0.99 & 3.41 & 0.87\\ 
		
		NB \cite{rish2001empirical} & 0.99 & 1.0 & 0.99 & 0.99 & 0.00065 & 0.99\\
		
		RF \cite{farnaazjabbar2016} & 0.99 & 0.99 & 1.0 & 0.99 & 1.61 & 0.87\\
		
		ANN \cite{saritas2019performance} & 0.99 & 0.99 & 1.0 & 0.99 & 5.30 & 0.5\\
		
		LR \cite{7060117} & 0.99 & 0.99 & 1.0 & 0.99 & 0.00030 & 0.56\\
		\bottomrule 
	\end{tabular}
	
\end{specialtable}

Table \ref{DURCON} shows the confusion matrix for NB which shows the disparity between the data in the dataset.

\begin{specialtable}[H]
	\caption{DoS UDP NB confusion matrix.}\label{DURCON}
		\setlength{\tabcolsep}{14.1mm}
	\begin{tabular}{ccc}
				\toprule 
		& \textbf{Actual Label} &\\ 
			\midrule 
		Predicted label & No attack & Attack\\
			\midrule 
		No attack & 8 & 0\\
		Attack & 4 & \hl{209703}\\
		\bottomrule 
	\end{tabular}
\end{specialtable}

Table \ref{DUWCR} shows the results of DoS UDP data with weighted classes enabled. ANN is unaffected and maintains poor log loss and ROC AUC scores. SVM has gained perfect precision but lost perfect recall with an increase in log loss and ROC AUC. DT has also swapped its precision and recall scores with an increase in both log loss and ROC AUC scores. RF has lost its perfect recall and increased its log loss and ROC AUC. LR has improved its log loss, ROC AUC, and gained perfect precision while losing perfect recall.
\begin{specialtable}[H]
	\caption{DoS UDP weighted classes results.}\label{DUWCR}
		\setlength{\tabcolsep}{1.95mm}
	\begin{tabular}{ccccccc}
		\toprule 
		\textbf{Algorithms Used} & \textbf{Accuracy }& \textbf{Precision} & \textbf{Recall} & \textbf{F1 Score} & \textbf{Log Loss} & \textbf{ROC AUC}\\
		\midrule 
		KNN \cite{harrison2019} & n/a & n/a & n/a & n/a & n/a & n/a\\
		
			SVM \cite{noble2006support} & 0.99 & 1.0 & 0.99 & 0.99 & 0.00053 & 0.99\\
		
		DT \cite{sharmakumar2016} & 0.99 & 1.0 & 0.99 & 0.99 & 5.24 & 0.99\\ 
		
		NB \cite{rish2001empirical} & n/a & n/a & n/a & n/a & n/a & n/a\\
		
		RF \cite{farnaazjabbar2016} & 0.99 & 0.99 & 0.99 & 0.99 & 1.34 & 0.93\\
		
		ANN \cite{saritas2019performance} & 0.99 & 0.99 & 1.0 & 0.99 & 5.30 & 0.5\\
		
		LR \cite{7060117} & 0.99 & 1.0 & 0.99 & 0.99 & 0.00079 & 0.99\\
		\bottomrule 
	\end{tabular}
	
\end{specialtable}

Without weighted classes, NB is the best model having perfect precision with a low log loss and high ROC AUC. With weighted classes, both SVM and LR perform very well but SVM is the better model as it has the lower log loss of the two models.

\subsubsection{Model Comparison}

Table \ref{BESTMODEL} shows the best models for each of the datasets including both (with and without weighted classes). DT and RF are the models that appear the most in the table with ANN appearing frequently in the weighted classes column. Without the use of weighted classes, RF achieves the best performance. With weighted classes, ANN achieves the best performances. However, using weighted classes generally decreases the overall performance of the model.
\begin{specialtable}[H]
	\caption{Model comparison.}\label{BESTMODEL}
		\setlength{\tabcolsep}{8.1mm}
\begin{tabular}{ccc}
		\toprule 
		\textbf{Dataset} &\multicolumn{2}{c}{\textbf{Best Model}}\\
		&\textbf{No Weighted Classes} & \textbf{Weighted Classes}\\
		
		\midrule 
		Data Exfiltrantion &RF&RF\\
		
		DDoS HTTP &DT&ANN\\
		
		DDoS TCP &RF&RF\\
		
		DDoS UDP &KNN&RF\\
		
		Keylogging &DT&DT\\
		
		OS Scan &RF&ANN\\
		
		Service scan &RF&ANN\\
		
		DoS HTTP &DT&ANN\\
		
		DoS TCP &KNN&DT\\
		
		DoS UDP &NB&SVM\\
		\midrule
		Most Occurrences &RF&ANN\\
		
		\bottomrule 
	\end{tabular}
	
\end{specialtable}

\subsubsection{Multiclass Classification}

Table \ref{MR} shows the results for multi-class classification, KNN has the best performance metrics including the lowest log loss and the highest CKS. LR is the worst model with the lowest metrics including the lowest CKS and a high log loss beat only by SVM.

\begin{specialtable} [H]
	\caption{Multi-class results.}\label{MR}
		\setlength{\tabcolsep}{2.63mm}
	\begin{tabular}{ccccccc}
		\toprule 
		\textbf{Algorithms Used} & \textbf{Accuracy }& \textbf{Precision} & \textbf{Recall} & \textbf{F1 Score} & \textbf{Log Loss} & \textbf{CKS}\\
		\midrule 
		KNN \cite{harrison2019} & 0.99 & 0.99 & 0.99 & 0.99 & 0.042 & 0.99\\
		
			SVM \cite{noble2006support} &0.79 &0.79 &0.79 &0.79 &0.65 &0.75 \\
		
		DT \cite{sharmakumar2016} & 0.96 & 0.96 & 0.96 & 0.96 & 0.11 & 0.95\\ 
		
		NB \cite{rish2001empirical} & 0.94 & 0.94 & 0.94 & 0.94 & 0.30 & 0.93\\
		
		RF \cite{farnaazjabbar2016} & 0.95 & 0.95 & 0.95 & 0.95 & 0.30 & 0.94\\
		
		ANN \cite{saritas2019performance} & 0.97 & 0.97 & 0.97 & 0.97 & 0.066 & 0.97\\
		
		LR \cite{7060117} & 0.74 & 0.74 & 0.74 & 0.74 & 0.63 & 0.68\\
		\bottomrule 
	\end{tabular}
\end{specialtable}

Table \ref{MWCR} shows the results with weighted classes. KNN and NB cannot use weighted classes and SVM was not tested because of its excessively long running time. Weighted classes have reduced the performance metrics for all models apart from ANN, which has had a small decrease in log loss, making it the best model with weighted classes. 

\begin{specialtable} [H]
	\caption{Multi-class weighted classes results.}\label{MWCR}
			\setlength{\tabcolsep}{2.63mm}
	\begin{tabular}{ccccccc}
		\toprule 
		\textbf{Algorithms Used} & \textbf{Accuracy }& \textbf{Precision} & \textbf{Recall} & \textbf{F1 Score} & \textbf{Log Loss} & \textbf{CKS}\\
		\midrule 
		KNN \cite{harrison2019} & n/a & n/a & n/a & n/a & n/a & n/a\\
		
			SVM \cite{noble2006support} & n/a & n/a & n/a & n/a & n/a & n/a\\
		
		DT \cite{sharmakumar2016} & 0.92 & 0.92 & 0.92 & 0.92 & 0.46 & 0.90\\ 
		
		NB \cite{rish2001empirical} & n/a & n/a & n/a & n/a & n/a & n/a\\
		
		RF \cite{farnaazjabbar2016} & 0.86 & 0.86 & 0.86 & 0.86 & 0.79 & 0.83\\
		
		ANN \cite{saritas2019performance} & 0.97 & 0.97 & 0.97 & 0.97 & 0.063 & 0.97\\
		
		LR \cite{7060117} & 0.69 & 0.69 & 0.69 & 0.69 & 0.75 & 0.63\\
		\bottomrule 
	\end{tabular}
	
\end{specialtable}

Table \ref{KNNCON} shows that KNN performs very well with the multi-class dataset with all the classes having low amounts of incorrectly classified data.
\begin{specialtable} [H]
	\caption{KNN confusion matrix.}\label{KNNCON}
			\setlength{\tabcolsep}{2.28mm}
\begin{tabular}{cccccccccccc}
		\toprule
		& & & & &\multicolumn{3}{c}{\textbf{Predicted}}& & & & \\
			\midrule 
		True & 0 & 1 & 2 & 3 & 4 & 5 & 6 & 7 & 8 & 9 & 10 \\
		 
		0 &\cellcolor{blue!25}172 &0 &0 &2 &0 &1 &107 &50 &0 &2 & 1 \\
		
		1 & 1&\cellcolor{blue!25}4 & 0&0 &0 &2 &0 &0 &0 &0 &0 \\
		
		2 &0 &0 &\cellcolor{blue!25} 965& 4& 2& 0& 0& 0& 43& 1& 0 \\
		
		3 &0 &0 &1 &\cellcolor{blue!25}\hl{56368} & 3& 0& 0& 0& 0& 2& 3 \\
		
		4 &0 & 0& 0&4 &\cellcolor{blue!25}55296 & 0& 0& 0& 0& 0& 2 \\
		
		5 &0 &1 &0 &0 &0 &\cellcolor{blue!25}80 & 0& 0& 0& 0& 0 \\
		
		6 &48 &0 &0 &0 &0 &0 &\cellcolor{blue!25}18294 &565 &0 &0 &0 \\
		
		7 &12 &0 &0 &0 &0 &0 &395 &\cellcolor{blue!25}55357 &0 &0 &0 \\
		
		8 &0 &0 &38 &6 &3 &0 &0 &0 &\cellcolor{blue!25}1427 &1 & 0 \\
		
		9 &0 &0 &2 &9 &1 &0 &0 &0 &4 &\cellcolor{blue!25}55218 &2 \\
		
		10 & 0& 0& 0& 4& 1& 0&0 &0 &0 &1 &\cellcolor{blue!25}55496 \\
			&&&&&&&&&\\[-3ex]\bottomrule 
		
	\end{tabular} 
\end{specialtable}

Table \ref{SVMCON} shows that SVM performs poorly with the multi-class dataset with data exfiltration (1), DDoS HTTP (2), and key logging (5) data all being incorrectly classified. These classes are ones featuring low amounts of data, which could be the reason for the low accuracy.

\begin{specialtable} [H]
	\caption{SVM confusion matrix.}\label{SVMCON}
		\setlength{\tabcolsep}{2.755mm}
\begin{tabular}{cccccccccccc}
			\toprule
		& & & & &\multicolumn{3}{c}{\textbf{Predicted}}& & & & \\
		\midrule
		True & 0 & 1 & 2 & 3 & 4 & 5 & 6 & 7 & 8 & 9 & 10 \\
			
		0 &\cellcolor{blue!25}10 &0 &0 &2 &3 &0 &183 &111 &6 &15 &5 \\
		
		1 &0 &\cellcolor{blue!25}0 &0 &0 &4 &0 &0 &0 &2 &1 &0 \\
		
		2 &0 &0 &\cellcolor{blue!25}0 &296 &6 &0 &0 &0 &79 &630 &4 \\
		
		3 &0 &0 &0 &\cellcolor{blue!25}\hl{19626} &17561 &0 &0 &0 &55 &17778 &1357 \\
		
		4 &0 &0 &0 &429 &\cellcolor{blue!25}54506 &0 &0 &0 &0 &2 &365 \\
		
		5 &0 &0 &0 &0 &7 &\cellcolor{blue!25}0 &0 &0 &72 &2 &0 \\
		
		6 &0 &0 &0 &1 &0 &0 &\cellcolor{blue!25}13056 &5779 &1 &41 &29 \\
		
		7 &0 &0 &0 &1 &0 &0 &3097 &\cellcolor{blue!25}52658 &0 &8 &0 \\
		
		8 &0 &0 &0 &512 &17 &0 &0 &0 &\cellcolor{blue!25}56 &885 &5 \\
		
		9 &0 &0 &0 &1804 &442 &0 &0 &0 &48 &\cellcolor{blue!25}52933 &9 \\
		
		10 &0 &0 &0 &4021 &5139 &0 &0 &0 &0 &22 &\cellcolor{blue!25}46319 \\
		&&&&&&&&&\\[-3ex]
			\bottomrule
		
	\end{tabular}
\end{specialtable}

Table \ref{DTCON} shows the confusion matrix for DT multi-class classification. It can be observed that the model performs very well; however, the model appears to have difficultly in correctly classifying the data that  are  imbalanced in the  dataset. This is evident in Table \ref{DTCON} with data exfiltration (1), DDoS HTTP (2), and key logging (5)  being  incorrectly classified.
\begin{specialtable} [H]
	\caption{DT confusion matrix}\label{DTCON}
		\setlength{\tabcolsep}{2.7mm}
	\begin{tabular}{cccccccccccc}
		
		\toprule
		& & & & &\multicolumn{3}{c}{\textbf{Predicted}}& & & & \\
		\midrule
		True & 0 & 1 & 2 & 3 & 4 & 5 & 6 & 7 & 8 & 9 & 10 \\
		
		0 &\cellcolor{blue!25}3 &0 &0 &1 &20 &0 &98 &227 &0 &1 &4 \\
		
		1 & 0&\cellcolor{blue!25}0 & 0& 0&3 & 0& 0& 0& 0& 0& 0 \\
		
		2 &0 &0 &\cellcolor{blue!25}0 &0 &1084 &0 &0 &0 &0 &0 & 0 \\
		
		3 &0 &0 &0 &\cellcolor{blue!25}\hl{55648} & 0&0 &0 &0 &0 &0 &0 \\
		
		4 &0 &0 &0 &0 &\cellcolor{blue!25}55460 &0 &0 &0 &0 &0 &0 \\
		
		5 &0 &0 &0 &0 &84 &\cellcolor{blue!25} 0&0 &0 &0 &0 &0 \\
		
		6 &0 &0 &0 &0 &0 &0 &\cellcolor{blue!25}9620 &9474 &0 &0 &0 \\
		
		7 &0 &0 &0 &0 &0 &0 &0 &\cellcolor{blue!25}55504 & 0&0 &0 \\
		
		8 &0 &0 &0 &0 &0 &0 &0 &0 &\cellcolor{blue!25} 1597& 0& 0 \\
		
		9 &1 & 0& 0& 0& 0& 0& 0& 0& 0&\cellcolor{blue!25}55784 &0 \\
		
		10 &0 &0 &0 &0 &0 & 0& 0& 0& 0& 0&\cellcolor{blue!25}55387 \\
		&&&&&&&&&\\[-3ex]\bottomrule
		
	\end{tabular} 
\end{specialtable}

Table \ref{DTWCON} shows the confusion matrix for DT with weighted classes enabled. Using weighted classes has resulted in an overall decrease in the models performance, but has improved the correct classification of data for normal traffic (0), data exfiltration (1), and key logging (5). This has also resulted in DoS HTTP having all its data incorrectly classified.

\begin{specialtable} [H] 
	\caption{DT weighted classes confusion matrix.}\label{DTWCON}
			\setlength{\tabcolsep}{2.55mm}
	\begin{tabular}{cccccccccccc}
		
		\toprule
		& & & & &\multicolumn{3}{c}{\textbf{Predicted}}& & & & \\
		\midrule
		True & 0 & 1 & 2 & 3 & 4 & 5 & 6 & 7 & 8 & 9 & 10 \\
		
		0 &\cellcolor{blue!25}297 &2 &0 &1 &3 &10 &0 &21 &0 &10 &3 \\
		
		1 & 0&\cellcolor{blue!25}6 & 0& 0&3 & 0& 0& 0& 0& 0& 0 \\
		
		2 &0 &0 &\cellcolor{blue!25}0 &0 &1055 &0 &0 &0 &0 &0 & 0 \\
		
		3 &0 &0 &0 &\cellcolor{blue!25}\hl{55716} & 0&0 &0 &0 &0 &0 &0 \\
		
		4 &0 &0 &0 &0 &\cellcolor{blue!25}55324 &0 &0 &0 &0 &0 &0 \\
		
		5 &0 &0 &0 &0 &0 &\cellcolor{blue!25} 70&0 &0 &0 &0 &0 \\
		
		6 &1037 &0 &0 &0 &0 &0 &\cellcolor{blue!25}17965 &0 &0 &0 &0 \\
		
		7 &1442 &0 &0 &0 &0 &0 &0 &\cellcolor{blue!25}54099 & 0&0 &0 \\
		
		8 &0 &0 &0 &0 &0 &0 &0 &0 &\cellcolor{blue!25} 0& 0& 1520 \\
		
		9 &0 & 0& 0& 0& 0& 0& 0& 0& 0&\cellcolor{blue!25}55745 &0 \\
		
		10 &0 &0 &0 &0 &0 & 0& 0& 0& 0& 0&\cellcolor{blue!25}55674 \\
			&&&&&&&&&\\[-3ex]\bottomrule
		
	\end{tabular} 
\end{specialtable}

Table \ref{NBCON} shows the confusion matrix for NB multi-classification, which performs quite well with no classes having all the data incorrectly classified. The model is also able to handle the data disparity in the classes with the low data classes having good classification results.
\begin{specialtable} [H]
	\caption{NB confusion matrix.}\label{NBCON}
	\setlength{\tabcolsep}{2.26mm}
\begin{tabular}{cccccccccccc}
		\toprule
		& & & & &\multicolumn{3}{c}{\textbf{Predicted}}& & & & \\
		\midrule
		
		True & 0 & 1 & 2 & 3 & 4 & 5 & 6 & 7 & 8 & 9 & 10 \\
		
		0 &\cellcolor{blue!25}33 &1 &0 &1 &0 &9 &224 &62 &1 &5 &0 \\
		
		1 &0 &\cellcolor{blue!25}7 & 0&0 &0 &0 &0 &0 &0 &0 &0 \\
		
		2 &2 &0 &\cellcolor{blue!25}1008 &0 &0 &0 &0 &7 &0 &0 &0 \\
		
		3 &29 &0 &0 &\cellcolor{blue!25}\hl{56296} &0 &0 &0 & 29&23 &0 &0 \\
		
		4 &2 &0 &0 &0 &\cellcolor{blue!25}55298 & 0&0 &2 &0 &0 &0 \\
		
		5 & 0& 0& 0& 0& 0&\cellcolor{blue!25}81 &0 &0 &0 &0 &0 \\
		
		6 & 67& 0&0 &0 &0 &0 &\cellcolor{blue!25}18199 &641 &0 &0 &0 \\
		
		7 & 362&0 &0 &0 &0 &0 & 15754&\cellcolor{blue!25}39648 &0 &0 &0 \\
		
		8 & 0&0 &0 &0 &0 &0 &0 &0 &\cellcolor{blue!25}1475 &0 &0 \\
		
		9 &1 &0 &0 &0 &0 &0 &0 &55 & 0&\cellcolor{blue!25}55180 &0 \\
		
		10 &1 &0 &0 &0 &0 &0 &0 &1 &0 &0 &\cellcolor{blue!25}55499 \\
			&&&&&&&&&\\[-3ex]\bottomrule
		
	\end{tabular}
\end{specialtable}

Table \ref{RFCON} shows the results for RF multi-class classification, which has good classification accuracy for the classes that have lots of data. The classes with low data have no correctly classified data.
\begin{specialtable} [H]
	\caption{RF confusion matrix.}\label{RFCON}
	\setlength{\tabcolsep}{2.975mm}
\begin{tabular}{cccccccccccc}
		
		\toprule
		& & & & &\multicolumn{3}{c}{\textbf{Predicted}}& & & & \\
		\midrule
		True & 0 & 1 & 2 & 3 & 4 & 5 & 6 & 7 & 8 & 9 & 10 \\
		
		0 &\cellcolor{blue!25}0 & 0& 0& 16& 2& 0& 31& 100& 0& 172&14 \\
		
		1 &0 &\cellcolor{blue!25}0 &0 &5 &2 &0 &0 &0 &0 &0 &0 \\
		
		2 &0 &0 &\cellcolor{blue!25}0 &955 &20 &0 &0 &0 &0 &0 &0 \\
		
		3 &0 &0 &0 &\cellcolor{blue!25}\hl{56377} & 0& 0& 0& 0& 0& 0&0 \\
		
		4 &0 &0 &0 &95 &\cellcolor{blue!25}55207 &0 &0 &0 &0 &0 &0 \\
		
		5 &0 &0 &0 &77 &4 &\cellcolor{blue!25}0 &0 &0 &0 &0 &0 \\
		
		6 &0 &0 &0 &1 &0 &0 &\cellcolor{blue!25}9238 &9514 &0 &151 &3 \\
		
		7 &0 &0 &0 &0 &0 &0 &0 &\cellcolor{blue!25}55716 &0 &47 &1 \\
		
		8 &0 &0 &0 &1422 &0 &0 &0 &0 &\cellcolor{blue!25}0 &0 &53 \\
		
		9 &0 &0 &0 &0 &0 &0 &0 &0 &0 &\cellcolor{blue!25}55229 &7 \\
		
		10 &0 &0 &0 &10 &0 &0 &0 &0 &0 &0 &\cellcolor{blue!25}55491 \\
				&&&&&&&&&\\[-3ex]	\bottomrule
		
	\end{tabular}
\end{specialtable}

Table \ref{RFWCCON} shows the results of having weighted classes. It is shown that, despite the models having  lower correct classifications overall, \hl{they have}  performed better with low data and correctly classifying the classes.

\begin{specialtable} [H]
	\caption{RF weighted classes confusion matrix}\label{RFWCCON}
\begin{tabular}{cccccccccccc}
		
		\toprule
		& & & & &\multicolumn{3}{c}{\textbf{Predicted}}& & & & \\
		\midrule
		True & 0 & 1 & 2 & 3 & 4 & 5 & 6 & 7 & 8 & 9 & 10 \\
		
		0 &\cellcolor{blue!25}250 & 1& 1& 0& 1& 11& 8& 32& 10& 12&9 \\
		
		1 &0 &\cellcolor{blue!25}7 &0 &0 &0 &0 &0 &0 &0 &0 &0 \\
		
		2 &0 &0 &\cellcolor{blue!25}1013 &0 &0 &2 &0 &0 &0 &0 &0 \\
		
		3 &0 &0 &231 &\cellcolor{blue!25}\hl{37544} & 18602& 0& 0& 0& 0& 0&0 \\
		
		4 &0 &0 &0 &76 &\cellcolor{blue!25}55226 &0 &0 &0 &0 &0 &0 \\
		
		5 &0 &4 &1 &0 &0 &\cellcolor{blue!25}76 &0 &0 &0 &0 &0 \\
		
		6 &180 &11 &0 &0 &0 &0 &\cellcolor{blue!25}17748 &514 &443 &11 &0 \\
		
		7 &333 &0 &0 &0 &0 &0 &16281 &\cellcolor{blue!25}38936 &131 &83 &0 \\
		
		8 &0 &0 &0 &0 &0 &0 &2 &0 &\cellcolor{blue!25}1473 &0 &0 \\
		
		9 &3758 &0 &0 &0 &0 &0 &0 &0 &25 &\cellcolor{blue!25}50448 &1005 \\
		
		10 &1 &0 &0 &0 &0 &0 &0 &0 &0 &0 &\cellcolor{blue!25}55500 \\
			&&&&&&&&&\\[-3ex]	\bottomrule
		
	\end{tabular}
\end{specialtable}

Table \ref{ANNCON} shows the results for ANN multi-class classification. The model performs well except for exfiltration (1) and key logging (5), which have incorrectly classified data.
\begin{specialtable} [H]
	\caption{ANN confusion matrix.}\label{ANNCON}
	\setlength{\tabcolsep}{2.55mm}
\begin{tabular}{cccccccccccc}
		\toprule
		& & & & &\multicolumn{3}{c}{\textbf{Predicted}}& & & & \\
		\midrule
			
		True & 0 & 1 & 2 & 3 & 4 & 5 & 6 & 7 & 8 & 9 & 10 \\
		
		0 &\cellcolor{blue!25}4 &0 &1 &14 &0 &0 &219 &86 &4 &6 &1 \\
		
		1 &0 &\cellcolor{blue!25}0 &0 &7 &0 &0 &0 &0 &0 &0 &0 \\
		
		2 &0 &0 &\cellcolor{blue!25}981 &34 &0 &0 &0 &0 &0 &0 &0 \\
		
		3 &0 &0 &27 &\cellcolor{blue!25}\hl{56349} &1 &0 &0 &0 &0 &0 &0 \\
		
		4 &0 &0 &0 &4 &\cellcolor{blue!25}55298 &0 &0 &0 &0 &0 &0 \\
		
		5 &0 &0 &0 &81 &0 &\cellcolor{blue!25}0 &0 &0 &0 &0 &0 \\
		
		6 &0 &0 &0 &9 &0 &0 &\cellcolor{blue!25}15312 &3586 &0 &0 &0 \\
		
		7 &0 &0 &0 &0 &0 &0 &2701 &\cellcolor{blue!25}53063 & 0& 0& 0 \\
		
		8 &0 &0 &0 &59 &0 &0 &0 &0 &\cellcolor{blue!25}1415 &0 &1 \\
		
		9 &0 &0 &0 &55 &0 &0 &19 &0 &1 &\cellcolor{blue!25}55161 &0 \\
		
		10 &0 & 0& 0& 2& 0& 0& 0& 0& 0&0 &\cellcolor{blue!25}55499 \\
		&&&&&&&&&\\[-3ex]	\bottomrule
		
	\end{tabular}
\end{specialtable}

Table \ref{ANNWCCON} shows the results with weighted classes enabled. It is observable that the model is much better at classifying most classes with OS scan (6) and service scan (7) having the most incorrectly classified data. The models is also unable to correctly classify any data for normal data (0) and data exfiltration (1).

\begin{specialtable} [H]
	\caption{ANN weighted classes confusion matrix.}\label{ANNWCCON}
	\setlength{\tabcolsep}{2.41mm}
\begin{tabular}{cccccccccccc}
		\toprule
		& & & & &\multicolumn{3}{c}{\textbf{Predicted}}& & & & \\
		\midrule
		& & & & & \multicolumn{3}{c}{Predicted}& & & & \\
		True & 0 & 1 & 2 & 3 & 4 & 5 & 6 & 7 & 8 & 9 & 10 \\
		
		0 &\cellcolor{blue!25}0 &0 &1 &1 &0 &12 &228 &81 &4 &6 &2 \\
		
		1 &0 &\cellcolor{blue!25}0 &0 &0 &7 &0 &0 &0 &0 &0 &0 \\
		
		2 &0 &0 &\cellcolor{blue!25}1010 &0 &5 &0 &0 &0 &0 &0 &0 \\
		
		3 &0 &0 &0 &\cellcolor{blue!25}\hl{56377} &0 &0 &0 &0 &0 &0 &0 \\
		
		4 &1 &0 &0 &2 &\cellcolor{blue!25}55299 &0 &0 &0 &0 &0 &0 \\
		
		5 &0 &0 &0 &0 &0 &\cellcolor{blue!25}81 &0 &0 &0 &0 &0 \\
		
		6 &0 &0 &0 &0 &0 &0 &\cellcolor{blue!25}16950 &1956 &0 &1 &0 \\
		
		7 &0 &0 &0 &0 &0 &0 &4195 &\cellcolor{blue!25}51569 & 0& 0& 0 \\
		
		8 &0 &0 &0 &0 &0 &0 &0 &0 &\cellcolor{blue!25}1473 &0 &2 \\
		
		9 &0 &0 &0 &0 &0 &0 &0 &0 &4 &\cellcolor{blue!25}55232 &0 \\
		
		10 &0 & 0& 0& 0& 0& 0& 0& 0& 0&1 &\cellcolor{blue!25}55500 \\
		&&&&&&&&&\\[-3ex]\bottomrule
		
	\end{tabular}
\end{specialtable}
Table \ref{LRCON} shows the results for LR multi-class classification, which has poor performance overall with the low data classes and also having no correctly classified data.

\begin{specialtable} [H]
	\caption{LR confusion matrix.}\label{LRCON}
	\setlength{\tabcolsep}{2.58mm}
	\begin{tabular}{cccccccccccc}
		
		\toprule
		& & & & &\multicolumn{3}{c}{\textbf{Predicted}}& & & & \\
		\midrule
		True & 0 & 1 & 2 & 3 & 4 & 5 & 6 & 7 & 8 & 9 & 10 \\
		
		0 &\cellcolor{blue!25} 0&0 &0 &11 &2 &0 &200 &101 &0 &9 &12 \\
		
		1 &0 &\cellcolor{blue!25}0 &0 &7 &0 &0 &0 &0 &0 &0 &0 \\
		
		2 &0 &0 &\cellcolor{blue!25} 0& 93& 6& 0& 0& 0& 308&601 &7 \\
		
		3 &0 &0 &0 &\cellcolor{blue!25} 16470& 6901& 0& 0& 0& 44&\hl{24490} &8472 \\
		
		4 &0 &0 &0 &145 &\cellcolor{blue!25} 49111& 0& 0& 0& 0& 71&5947 \\
		
		5 &0 &0 &0 &81 &0 &\cellcolor{blue!25}0 &0 &0 &0 &0 &0 \\
		
		6 &2 &0 &0 &0 &0 &0 &\cellcolor{blue!25} 14690& 4186& 0& 0& 29 \\
		
		7 &0 &0 &0 &0 &0 &0 &3713 &\cellcolor{blue!25}52049 &0 &0 &2 \\
		
		8 &0 &0 &0 &139 &18 &0 &9 &0 &\cellcolor{blue!25}482 &819 &8 \\
		
		9 &0 &0 &0 &274 &453 &0 &2 &0 & 435&\cellcolor{blue!25}54035 &37 \\
		
		10 &0 &0 &0 &10658 &9195 & 0& 0& 0& 0&15&\cellcolor{blue!25}35633 \\
				&&&&&&&&&\\[-3ex]\bottomrule
		
	\end{tabular}
\end{specialtable}
Table \ref{LRWCCON} shows the results of having weighted classes. It is evident that the accuracy of overall classification has decreased;  however, the model shows improvement in classifying the low data classes.
\begin{specialtable} [H]
	\caption{LR weighted classes confusion matrix.}\label{LRWCCON}
	\setlength{\tabcolsep}{1.97mm}
	\begin{tabular}{cccccccccccc}
		
		\toprule
		& & & & &\multicolumn{3}{c}{\textbf{Predicted}}& & & & \\
		\midrule
		
		& & & & &\multicolumn{3}{c}{Predicted}& & & & \\
		True & 0 & 1 & 2 & 3 & 4 & 5 & 6 & 7 & 8 & 9 & 10 \\
		
		0 &\cellcolor{blue!25} 291&4 &0 &0 &2 &7 &10 &14 &0 &6 &1 \\
		
		1 &0 &\cellcolor{blue!25}7 &0 &0 &0 &0 &0 &0 &0 &0 &0 \\
		
		2 &0 &9 &\cellcolor{blue!25} 692& 0& 2& 0& 0& 0& 306&2 &4 \\
		
		3 &0 &105 &432 &\cellcolor{blue!25} \hl{14010}& 7516& 0& 0& 0& 1004&\hl{23636} &9674 \\
		
		4 &0 &44 &50 &186 &\cellcolor{blue!25} 52423& 0& 0& 0& 34& 71&2494 \\
		
		5 &0 &3 &0 &0 &0 &\cellcolor{blue!25}78 &0 &0 &0 &0 &0 \\
		
		6 &3570 &0 &0 &0 &0 &0 &\cellcolor{blue!25} 13753& 1582& 2& 0& 0 \\
		
		7 &4353 &0 &0 &0 &0 &0 &9652 &\cellcolor{blue!25}41758 &0 &0 &1 \\
		
		8 &0 &15 &710 &0 &8 &0 &0 &0 &\cellcolor{blue!25}736 &4 &2 \\
		
		9 &0 &0 &954 &424 &453 &0 &11 &0 & 2479&\cellcolor{blue!25}50907 &8 \\
		
		10 &0 &177 &197 &10008 &9850 & 0& 0& 0& 216&16&\cellcolor{blue!25}35037 \\
			&&&&&&&&&\\[-3ex]\bottomrule
		
	\end{tabular}
\end{specialtable}

\section{Conclusions}
\label{section4}
In this  paper, state-of-the-art ML algorithms  are  compared in terms of accuracy, precision, recall, F1 score, and log loss on both weighted and non-weighted Bot-IoT dataset. It  is shown that the performance of RF in terms of accuracy and precision is the best with the non-weighted dataset. However, in a weighted dataset, ANN has higher accuracy for binary classification. In multi-classification, KNN and ANN  are  highly accurate for weighted and non-weighted datasets, respectively. From the results, it is evident that, when all types of attack have weighted datasets, ANN predicts the type of attack with higher accuracy. 

In the future, we intend to adopt the models explored in this research into an IDS prototype for testing using diverse data including a mix of attacks to validate the multi-class functionality of models.

\authorcontributions{\hl{Conceptualization,} A.C.; Methodology, A.C.;
	Validation, A.C., J.A., and R.U.; Formal Analysis, A.C.; Investigation, A.C., J.A., R.U., and B.N.; resources, A.C.; writing---original draft preparation, A.C. and J.A.; writing---review and editing, A.C., J.A., R.U., B.N., F.M., S.U.R., F.A., and W.J.B.; supervision, J.A. and W.J.B.; and funding acquisition, F.A. and  J.A. All authors have read and agreed to the published version of the manuscript.}

\funding{\hl{text.}}
\institutionalreview{\hl{text.}}

\informedconsent{\hl{text.}}

\dataavailability{\hl{text.}} 
\conflictsofinterest{The authors declare no conflict of interest.}



\end{paracol}
\reftitle{References}

\end{document}